\documentclass[11pt, a4paper, logo, onecolumn, copyright]{googledeepmind}

\usepackage[authoryear, sort&compress, round]{natbib}
\usepackage{array}
\usepackage{multirow}
\usepackage{caption}
\usepackage{tikz}
\usetikzlibrary{arrows,positioning}
\usepackage[titles]{tocloft}
\usepackage{amsthm}
\usepackage{blindtext}
\usepackage{hyperref}

\theoremstyle{definition}

\theoremstyle{conjecture}
\newtheorem{conj}{Conjecture}[]

\author[1]{Logan Cross}
\author[1]{Jordi Grau-Moya}
\author[1,2]{William A. Cunningham}
\author[1]{Alexander Sasha Vezhnevets}
\author[1]{Joel Z.~Leibo}

\affil[1]{Google DeepMind}
\affil[2]{University of Toronto}

\usepackage{tcolorbox}
\usepackage{wrapfig}

\title{A Generative Model of Conspicuous Consumption and Status Signaling}

\reportnumber{} 

\renewcommand{\today}{}

\begin{document}

\begin{abstract}

Status signaling drives human behavior and the allocation of scarce resources such as mating opportunities, yet the generative mechanisms governing how specific goods, signals, or behaviors acquire prestige remain a puzzle. Classical frameworks, such as Costly Signaling Theory, treat preferences as fixed and struggle to explain how semiotic meaning changes based on context or drifts dynamically over time, occasionally reaching tipping points. In this work, we propose a computational theory of status grounded in the theory of appropriateness, positing that status symbols emerge endogenously through a feedback loop of social observation and predictive pattern completion. We validate this theory using simulations of groups of Large Language Model (LLM)-based agents in the Concordia framework. By experimentally manipulating social visibility within naturalistic agent daily routines, we demonstrate that social interactions transform functional demand into status-seeking behavior. We observe the emergence of price run-ups and positive price elasticity (Veblen effects) for both real-world luxury items and procedurally generated synthetic goods, ruling out pretraining bias as the sole driver. Furthermore, we demonstrate that "influencer" agents can drive the endogenous formation of distinct subcultures through targeted sanctioning, and find that similar social influence effects generalize to non-monetary signaling behaviors. This work provides a generative bridge between micro-level cognition and macro-level economic and sociological phenomena, offering a new methodology for forecasting how cultural conventions emerge from interaction.

\end{abstract}

\maketitle









\section{Introduction}

Status signaling fundamentally drives human behavior, wherein individuals take actions to advertise otherwise unobservable traits like wealth, fitness, or group allegiance. This drive is adaptive and evolutionarily conserved, as status is a relative ranking that mediates the allocation of scarce resources such as attention, allies, and mating opportunities \citep{zahavi1975mate,grafen1990biological}. Beyond resource allocation, these displays serve a semiotic function: they act as a system of social classification that allows individuals to distinguish in-group members and infer the latent characteristics of others \citep{simmel1904sociology, bourdieu1979distinction, jordan2016third, spence1978job, veblen1899theory}. Modern evolutionary accounts suggest that these signals also facilitate cultural learning by helping individuals identify prestigious models to imitate \citep{henrich2001evolution,henrich2015secret}.

Yet, the generative mechanism governing these displays remains a puzzle. Why do some goods, behaviors, or ideas become potent status symbols while others fade, and why do their meanings shift so dynamically across cultures and time \citep{barker2019content}? Researchers have struggled to empirically isolate the mechanisms, as these signals emerge organically and are socially constructed. Field data about the consumption of status goods confound prestige, marketing, and social diffusion \citep{manski1993identification,hartmann2008modeling}. Experiments in the lab struggle with ecological validity and making status and visibility realistic without losing control \citep{levitt2007laboratory,harrison2004field}. Understanding not just what functions as a signal but also how signals emerge, shift, and propagate is a central problem in social science.

Current theoretical frameworks offer only a patchwork of independent explanations for these phenomena. The rational-choice theory of human behavior struggles to explain why status goods frequently exhibit upward-sloping demand curves (Veblen effects) as the Law of Demand posits that demand falls when prices rise \citep{becker1976economic,varian2014intermediate,veblen1899theory,bagwell1996veblen}.
Costly Signaling Theory (CST) explains conspicuous consumption by treating expenditure as a credible signal of status because the cost is hard to fake for low status individuals \citep{zahavi1975mate,grafen1990biological}. However, even though CST explains why costly signals are honest and have stable equilibria, high production cost is neither necessary nor sufficient to create a status signal \citep{lachmann2001cost, barker2019content}. 

\begin{figure}
    \centering
    \includegraphics[width=0.85\textwidth]{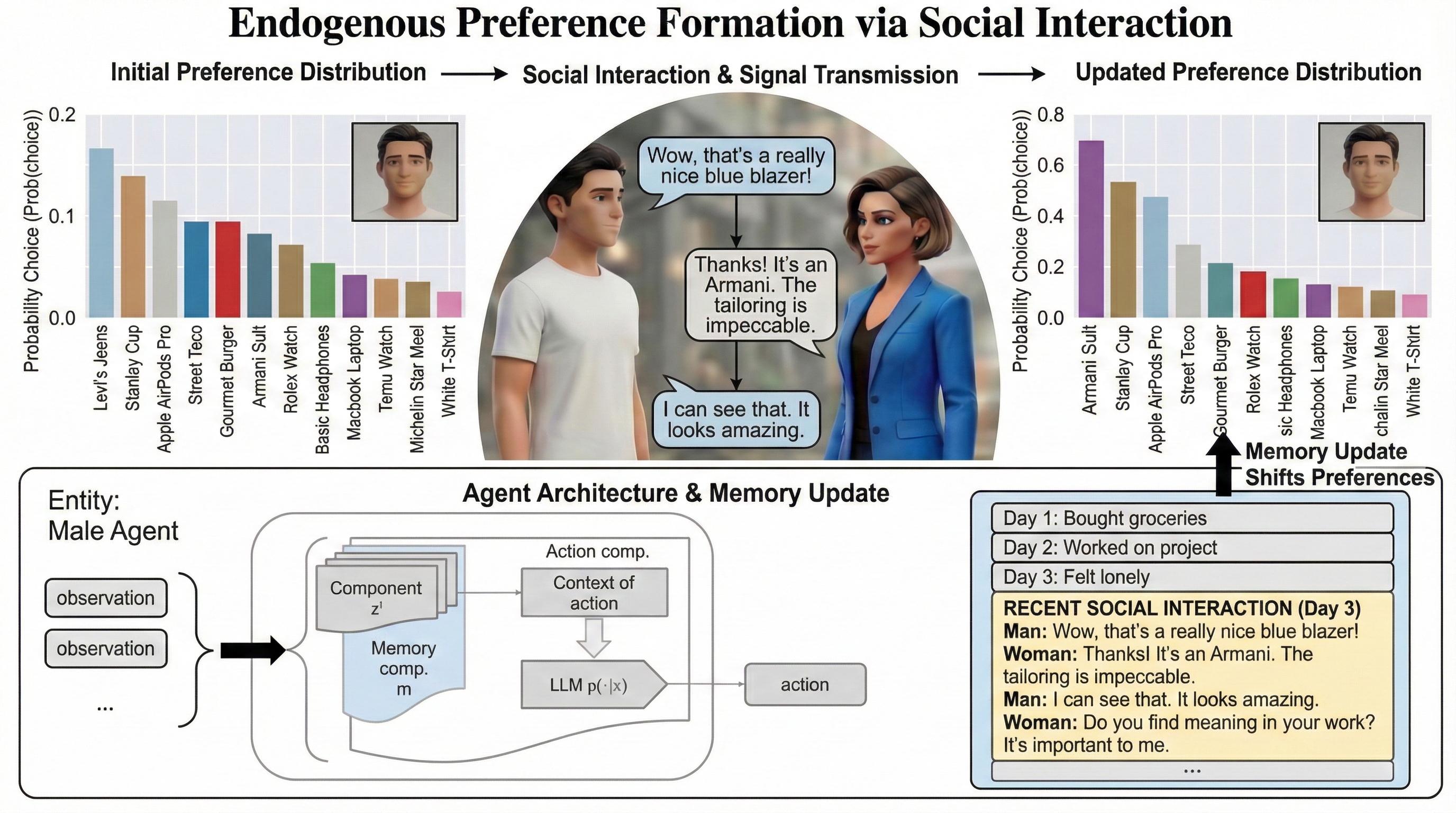}
    \caption{The figure illustrates how status preferences shift dynamically through agent-to-agent interaction rather than hard-coded utility functions. (Left) The Initial Preference Distribution shows an agent with a current snapshot of its preferences. (Middle) During the Social Interaction \& Signal Transmission phase, the focal agent (left) interacts with a partner displaying a distinct status signal (an Armani blazer). The dialogue, which includes both casual conversation and specific acknowledgment of the item, is encoded into the agent's memory stream (depicted in the architecture schematic below). (Right) The Updated Preference Distribution shifts toward the observed status symbol (Armani Suit).}
    \label{fig:endog}
\end{figure}

Research on neutral markers demonstrates that nearly costless behaviors can emerge as in-group signaling conventions. For example accents, slang, religious markings, conventional in-group badges, and moralized reposts of political news—are nearly costless to produce yet remain reliable social signals \citep{cohen2012evolution,samarasinghe2019evaluations,barker2019content,tibbetts2004socially,tajfel1971social,ghezae2024partisans}. These signals solve a coordination problem in which the value of a symbol emerges through equilibrium selection: once a community converges on an arbitrary marker, in-group members imitate that marker over alternatives \citep{lachmann2001cost,centola2018experimental,lewis1969convention,skyrms2010signals}. Social imitation of arbitrary symbols and rituals serves as a reliable learning heuristic as humans are embedded in complex environments where knowledge is opaque and culturally inherited \citep{henrich2015secret}. Moreover, high cost is not sufficient for status, as the semiotic value of a signal is often subject to rapid cultural drift. Expensive goods frequently lose their signaling power because they become accessible to the 'wrong' groups or fail to align with shifting aesthetic norms, rendering them passé despite their continued expense \citep{simmel1904sociology, leibenstein1950bandwagon, bourdieu1979distinction, pesendorfer1995design, berger2008drives}. 

Across these domains, rather than cost, the binding constraint is semiotic \textbf{appropriateness}—how observers interpret a signaling action given precedent, social role, context, and shared cultural expectations. Cost can contribute to perceived appropriateness—but so can aesthetic form, identity alignment, and the weight of prior examples. A unified model of signaling must therefore explain why costly signals sometimes work, why cheap signals often work, and why entire cultural regimes of signaling can shift rapidly as conventions change. We propose such a model unified model grounded in the theory of appropriateness, positing that status symbols emerge endogenously through a feedback loop of social observation and imitation via predictive pattern completion \citep{bowles1998endogenous, leibo2024theory, leibo2025norms}. Predictive pattern completion conceptualizes humans as culturally trained generative models who produce actions that are appropriate for their context by developing a habitus — an intuitive sense for the “rules of the game” coined by sociologist Pierre Bourdieu \citep{bourdieu1979distinction}. In this view, signals acquire value through social diffusion: individuals observe peers, update expectations about what "people like us" do, and imitate the signaling behaviors that align with that shared context.



We operationalize our theory using multi-agent societies of generative agents in Concordia \citep{park2023generative,vezhnevets2023generative,vezhnevets2025multi,anthis2025llm}. As previous work validated that these Large Language Model (LLM)-based agents can reproduce human signaling dynamics in controlled experiments \citep{cross2025validating}, we scale these mechanisms to a naturalistic setting to observe the endogenous emergence of status signals. This approach allows us to create ecologically-valid yet precisely controllable social environments, where we can isolate the essential ingredients for status dynamics to emerge without identification problems \citep{manski1993identification}. It enables us to manipulate who talks to whom, what is publicly visible, and how choices are made, while measuring the emergent outcomes and revealed preferences of the agents \citep{samuelson1938note}.

In our experiments, LLM agents live daily lives consistent with synthetic personas by observing daily events and making decisions that have private and social effects. Agents can buy food, gadgets, clothing, and accessories in a marketplace. The agents have social interactions---first dates---where their visible outfit/accessories are sampled from inventories, and they chat freely about their lives. With this social layer turned on, we observe high demand and price run-ups for status goods (e.g., Chanel handbags, Rolex watches) and for a toy collectible (Labubu) that is currently trending internationally but had not been well known prior to the base model’s knowledge cutoff \citep{yang2025study,lawrence2025scarcity}. The hype for Labubu was so high that we observed positive price elasticity over time, meaning Labubu became a Veblen good. Without the social layer, prices remain static and agents purchase status goods at a significantly lower rate. We reproduced this effect with synthetically generated status goods, ruling out the possibility that these results are driven by the brand prestige ingrained in the LLMs prior. Observing these price-runs and Veblen effects in our "field data" presents a identification problem: disentangling whether a high price causes high demand (per Costly Signaling Theory) or if socially constructed demand drives up the price through extensive hype. Our fixed-price control resolves this ambiguity as status purchases increased even more with fixed initial prices, demonstrating that social contagion—not signal cost—causally drives consumption.

To test the generality of this social mechanism, we extended our framework to reproduce human examples of non-monetary signaling that are documented in the literature using procedurally generated status signaling simulations \citep{jordan2020signaling,glazer1996signaling,barker2019content,skyrms2010signals}. Given a trait to signal such as group identity, political allegiance, or altruism, we synthetically generate diverse scenarios that evaluate whether the agents prefer to produce a behavior that emits a social signal or a private behavior. Similarly to the marketplace simulations, signaling behaviors increased in conditions where agents socially interact and are exposed to these signals from others. When we hold everything else constant and remove the social life, the desire for these status symbols reliably decreases.

This research therefore makes three core contributions. First, we offer a new computational theory of status signaling grounded in the theory of appropriateness that moves beyond static utility functions and costly-signaling models. Second, we introduce a novel methodology using generative agent societies to directly simulate endogenous preference formation (the process where desires are shaped by social interaction, rather than being fixed or innate) \citep{bowles1998endogenous}, showing in our experiments that social visibility is the critical ingredient that transforms functional demand into status-seeking and Veblen effects. Finally, we deliver a generalizable, open-source framework that serves as a computational theory apparatus, allowing social scientists to generate and test falsifiable predictions about how cultural conventions emerge from micro-level interactions \footnote{Code is available at: \url{https://github.com/google-deepmind/concordia/tree/main/examples/signaling}}. We can now move from observing social conventions and norms to computationally generating them, offering a new way to generate testable predictions about real-world phenomena such as how viral trends emerge, how consumer tastes shift, and how communities develop their own unique status symbols and semiotic codes.

\begin{figure}
    \centering
    \includegraphics[width=0.9\textwidth]{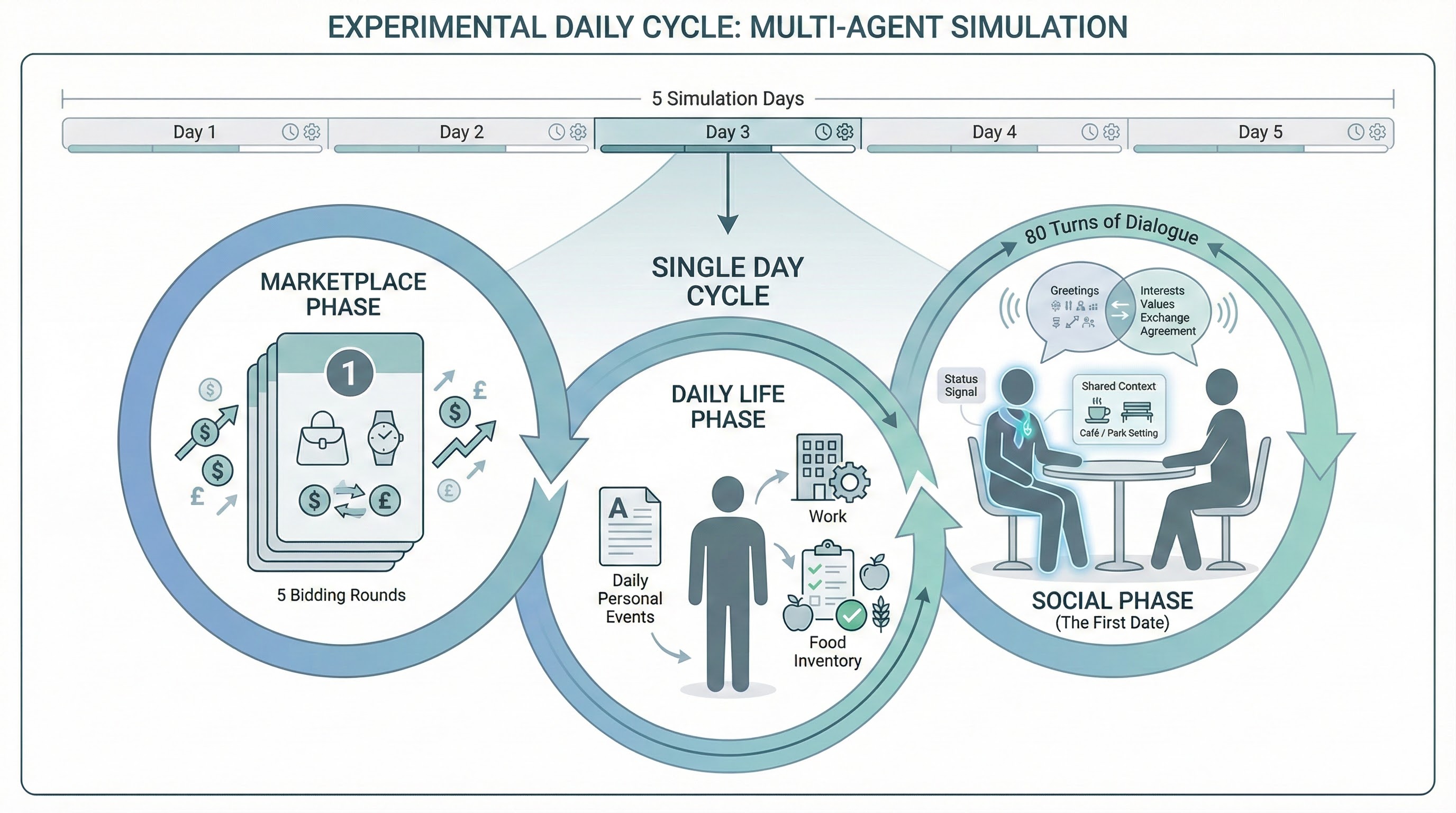}
    \caption{\textbf{Simulation Timeline and Daily Phases.} The experiment runs for 5 days. The inset details the daily workflow: agents first compete for resources in 5 rounds of marketplace activity , followed by private daily life events. Finally, agents enter the social phase, where they engage in "first date" interactions consisting of visual observations of status goods and 80 turns of conversation.}
    \label{fig:cycle}
\end{figure}


\section{Theoretical Framework: Predictive Pattern Completion as the Generative Mechanism of Status Signaling}

Status signalling is a dynamic process of social communication and cultural construction. We define a signaling behavior as an action that serves as an intersubjective symbol, allowing observers to infer latent, desirable traits from the signal. Through a cycle of imitation and peer sanctioning, humans learn to produce symbols in contexts where they have previously observed them, eventually internalizing these patterns as automatic habits that sometimes persist even in private \citep{jordan2020signaling,leibo2024theory,leibo2025norms}. As this process spreads, social contagion leads to a co-constitution of a convention or social norm. Over time in a culture, these accumulated signals coalesce into a dynamic semiotic code—a complex symbolic language that allows individuals to implicitly broadcast their group identity and social standing.

What makes a good signal? From a classical view, evolutionary pressures select a status signal to pragmatically assist cooperation and social interaction \citep{zahavi1975mate,grafen1990biological,gintis2001costly,alexander1987biology, boyd1989evolution, nowak1998evolution,young2015evolution, henrich2009evolution}. Signals help produce stable cooperative equilibria by enabling reputation systems that efficiently mediate the allocation of scarce resources \citep{jordan2023pull}. Partner choice is the canonical example of these status markets. In evolutionary biology, the peacock's tail is a costly, wasteful handicap that honestly signals genetic fitness to a potential mate \citep{zahavi1975mate}. In human contexts, conspicuous consumption, such as buying expensive luxury cars or clothing, can function as a similar sexual signaling system \citep{miller2001mating}. Our simulation's "first date" context is a direct implementation of this "partner choice" motivation. This classical, Costly Signaling Theory (CST), formalized by theorists from Zahavi to Gintis, Smith, and Bowles, posits that scarcity and cost certify underlying value \citep{gintis2001costly,zahavi1975mate,henrich2009evolution}. CST poses that for a signal to be honest and evolutionarily stable, it must be hard to fake without the underlying trait it signals, such as wealth, fitness, or religious belief, giving it credibility. This cost is often non-monetary; for instance, the time and intellectual effort required to receive an advanced degree provides a reliable signal of ability to employers \citep{spence1978job}. Similarly, Credibility Enhancing Displays (CREDs) such as martyrdom, celibacy, or painful rituals serve as costly signals by providing hard-to-fake evidence of a person's genuine commitment to a religious or cultural belief \citep{henrich2009evolution}. Likewise, participation in political protests entails safety and social risks that serve as credible signals of conviction and commitment to the political in-group \citep{lohmann1993signaling}.

In psychology experiments, Jordan and colleagues demonstrate that individuals use costly signals, such as third-party punishment (TPP) of norm violators, to signal their own unobservable trustworthiness \citep{jordan2016third}. Evaluators, in turn, use these signals to select partners for cooperation. Since the signal was costly, evaluators can infer that the signal was honest and reliable. A critical finding from this research is that signals are not static; they are highly context-sensitive performances \citep{jordan2016third,jordan2023pull}. The TPP signal attenuates when a more informative signal, such as the opportunity to directly help someone, becomes available. Individuals strategically switch to the signal that is most informative of their unobserved trait for the given context.

This context-sensitivity helps redefine the role of "cost," suggesting a more refined, relational view. What truly matters is the semiotic meaning of the symbol from the observer's perspective, for which production cost is only one potential component. The costly signaling framework is tied to a rational-actor theory of behavior \citep{becker1976economic,varian2014intermediate}. It casts the sender and observer as a "rational" payoff maximizing actors responding to fixed incentives. In this model, the observer wants to cooperate with a high value sender to increase their payoff after performing a specific, logical inference: they perceive a costly signal as honestly identifying high value types because it would be metabolically or financially difficult to fake for low value types. Additionally, the sender is strategically motivated to choose a costly signal based on calculating that the subsequent social benefits will outweigh the cost. However, this rational-actor view is incomplete and suffers from being context-insensitive, static, and path independent. It struggles to explain why a culture chooses one of many possible costly signals or why costless signals are often used, because it treats the cost-to-signal effectiveness link as a universal rule. For instance, empirical surveys of human signaling reveal that "materially cheap" neutral markers—such as etiquette or donning team colors—often function as highly effective signals of commitment to the in-group despite lacking intrinsic waste \citep{lachmann2001cost,barker2019content,cohen2012evolution,ghezae2024partisans,samarasinghe2019evaluations,tajfel1971social,tibbetts2004socially,mcelreath2003shared}. In these contexts, reliability is not enforced by production costs, but by "audience-dependent" risks: the reputational damage and social sanctioning imposed on cheaters who are discovered faking the signal. Thus, the "cost" in these behaviors is not a fixed attribute of the signal itself, but an emergent property of the specific normative climate that polices it.

Additionally, CST cannot fully explain the dynamic nature of signals that wax and wane in use (fashion trends and memes for example). The mathematical models of CST often fundamentally treat the signal's payoffs as a static, exogenous assumption, usually operationalized by a set of utility functions or payoff matrices pre-designed by the modeler, rather than endogenously generated preferences that emerge from the model simulation itself \citep{bowles1998endogenous}. While more sophisticated modeling frameworks like \citet{pesendorfer1995design} successfully formalize fashion cycles as a CST signaling game, they primarily address the decay of a signal rather than its genesis or content. In these models, agents rationally abandon a trend not because their tastes have changed, but because the signal’s informational value has degraded due to low status individuals adopting the trend. However, this view still treats the initial coordination on a specific signal as a given exogenous variable rather than an endogenous property that is continuously constructed. It leaves unanswered the problem of equilibrium selection: among the infinite set of possible costly behaviors, why does a population converge on a specific arbitrary token—like a Labubu doll—when many alternatives possess similar cost structures? By collapsing the multidimensional semiotic meaning of a signal into a single cost variable, CST fails to explain the diversity of highly contextual signaling behaviors and the reality of socially constructed preferences. The tenets of CST have been useful for explaining why a costly signal is stable once established, but lack a generative mechanism to explain how agents discover and select the signal in the first place.

A similar limitation applies to evolutionary models of indirect reciprocity and reputation \citep{alexander1987biology, boyd1989evolution, nowak1998evolution, panchanathan2004indirect,jordan2023pull}. While these frameworks extend CST beyond dyadic interactions to explain how broader community norms are enforced, they also treat the content of the norm as an exogenous fixed parameter. To illustrate this, consider the arbitrary convention of wearing a necktie to signal professionalism. These models excel at demonstrating why, once such a convention or norm is established, an individual is strategically incentivized to wear the useless strip of cloth to maintain their reputation and avoid social sanctioning. However, because the behavior is hard-coded as "good" in the model's assumptions, it cannot explain why the signal converged on a necktie rather than a sash, a specific color, or a lapel pin. By strictly separating the strategic incentives from the cultural construction of the signal, these models miss the feedback loop where the `rules of the game' are themselves continuously rewritten by the players.



Our framework, in contrast, aligns with a theory of appropriateness that models humans not as a rational actors maximizing utility, but as culturally trained generative models \citep{leibo2024theory,leibo2025norms}. These generative models act via predictive pattern completion: they take in context variables and sample from a predictive distribution of likely actions that are appropriate for that context given the cultural conventions and norms they observed in their training data and in-context memory. By observing other agents in their culture, the generative models continuously update their priors of what's appropriate and imitate others' behaviors and desires \citep{girard1965deceit}. The sociological framework of Pierre Bourdieu, for instance, highlights this different cognitive pathway via the concept of \textit{habitus} \citep{bourdieu1979distinction}. Habitus is the set of internalized dispositions, the embodied "feel for the game", that allows an agent to instinctively track and produce signals that are perceived as appropriate or `distinguished' in a given social context. Here, cultural capital functions as the reputation system, acquired by the education and upbringing necessary to know what to signal to demonstrate commitment to your in-group. The signals classify and distinguish us from others. For example, the bourgeois class in Bourdieu's time demonstrated their "taste of luxury" by prioritizing form and manner over the "taste of necessity" of the working class, which values substance and utility. This translated into preferring the presentation of food over how "filling" or "economical" it was, and favoring stylized, formal art over more "vulgar" forms of enjoyment. Veblen's \textit{The Theory of the Leisure Class} similarly describes how in the highest social strata, hard labor is perceived as a sign of economic weakness, such that non-productive uses of time (ie. learning dead languages like Latin and Greek, practicing etiquette) are a status symbol. In contrast, modern life has demonstrated a Veblenian inversion, as Western culture revalorizes working long hours, transforming what once marked subordination into a badge of agency and ambition \citep{bellezza2017conspicuous}. The fact that high cost can signal both low status (labor) and high status (grind culture) depending on the context is a significant flaw for simple, costly signaling models. In our refined framework, absolute "cost" is not the master variable; it is merely one of several factors—such as aesthetic form, differentiation, and social convention—that moderates a signal's effectiveness by shaping its ultimate semiotic meaning for the observer. Thus, the observer of an effective signal is not just calculating production cost; they are recognizing a set of behaviors that signal a shared habitus. 

Evolutionary anthropology provides robust empirical evidence for this learning mechanism of social imitation. \citet{henrich2015secret} synthesizes extensive research demonstrating that humans evolved primarily as adaptive cultural learners who blindly copy the behaviors of role models. Because the true mechanisms of evolutionary success are causally opaque and our cognition is bounded, evaluating the intrinsic utility of every possible action is computationally intractable. Instead, we rely on efficient heuristics like prestige-biased transmission, imitating the behaviors of high status individuals \citep{henrich2001evolution}. This instinct is vividly captured in comparative "over-imitation" experiments: whereas chimpanzees pragmatically skip causally useless steps to open a puzzle box, human children faithfully reproduce every arbitrary, wasteful motion demonstrated by an adult \citep{horner2005causal}. Henrich additionally illustrates this with anthropological case studies, such as the grueling, multi-step traditions indigenous Amazonians use to detoxify bitter manioc. Because the chronic cyanide poisoning caused by skipping a seemingly useless washing step takes years to manifest, individuals cannot safely optimize the recipe through rational trial-and-error; survival requires blindly imitating the inherited cultural convention. By copying a prestigious model's entire suite of behaviors—including seemingly arbitrary behaviors and fashions—we bypass the need for perfect information, adopting conventions simply because "people like us" do them. Within our computational framework, this cultural learning serves as the cognitive engine for acquiring \textit{habitus}.

Here, `arbitrary' does not mean random; rather, it denotes that these conventions are historically contingent and heavily path-dependent \citep{young2015evolution,lewis1969convention,skyrms2010signals}. Theorists emphasize that this arbitrariness—the capacity for populations to lock into any one of multiple stable equilibria—is the mechanism that generates and sustains the vast diversity of distinct cultural norms and subcultures observed globally \citep{boyd1992punishment,mcelreath2003shared,leibo2025societal}. As these local equilibria stabilize, they coalesce into "thick" semiotic codes—deeply embedded, historically grounded systems of meaning that enforce in-group solidarity, rather than "thin," universally applicable principles \citep{walzer2019thick}. Ultimately, the social construction of status signals is the product of two interacting dynamics across long and short timescales: the deeply ingrained, "thick" cultural priors that supply the inherited meaning of a behavior, and the social influence and imitation that drives its contagion.

Modeling these diverse signaling dynamics requires a mechanism that can generate, propagate, and update expectations about thick social meaning. The theory of appropriateness provides such a mechanism, which posits that individual behavior is generated via predictive pattern completion and can be modeled with LLM agents \citep{leibo2024theory,leibo2025norms,vezhnevets2023generative}. While traditional signaling models struggle to capture this high-dimensional semantic complexity, LLMs are uniquely suited to operationalize models with such profound context dependence \citep{kommers2025meaning}. By conditioning their predictive distributions on rich, natural-language context variables, LLMs can directly process and generate the "thick descriptions" required to navigate nuanced social realities.

This brings us to our paper's core theoretical claim. We posit that status signaling behaviors are culturally learned conventions, not manifestations of individual utility functions. We define agents as "convention-sensitive" if their behavior changes as a function of the "weight of precedent"—a generative mechanism where the accumulation of memories regarding others' past actions acts as training data that biases current decision-making. We formalize our central hypothesis as follows:

\begin{conj}[Social Exposure Increases Signaling through the Weight of Precedent]
Agents are convention-sensitive. Let $a_s$ be an action of status signaling behavior and $a_{ns}$ be any alternative action (e.g., purchasing a functional good or private behavior). Let $p(a_s | c)$ be the baseline probability of status signaling behavior in context $c$ with no social memories. Social exposure (observing others' behaviors, possessions, and conversations) provides new memories $m_s$ that increase the counterfactual weight of precedent for $a_s$. These new memories increase the influence of past precedents for the action, making it significantly more likely to be chosen compared to a counterfactual situation where the agent had no such memories. Therefore, we predict that the probability of the status signaling behavior will be significantly higher in conditions with social exposure of the signal: $p(a_s | c, m_s) > p(a_s | c)$.
\end{conj}

Our design, which manipulates the presence of social exposure of a status symbol in memory ($m_s$), is built to test this conjecture. We run simulations of small generative agent societies and experimentally modify whether agents interact with each other and can view the public signals they emit as a consequence of their actions. In our generative-agent environment, public exposure of the signal and conversation about it acts as the training data for pattern completion and imitation with in-context learning. In Part 1, we afford agents the opportunity to purchase functional and status goods in a marketplace. In the Social condition, agents periodically go on first dates and encounter the clothing or accessories their date purchased, allowing the agents to interpolate shared representations of what is desirable, causing price run-ups for the status goods they observed. We repeat this for synthetically generated goods that are not baked into the prior of the LLM as high status, constructing new status symbols from scratch in silico. This demonstrates that once an internalized schema about status symbols is learned, arbitrary tokens can be filled in and re-signified as the next brand or social convention without centralized coordination. In Part 2, we generalize this social contagion process to non-monetary domains, demonstrating that many arbitrary behaviors can be reinforced through the weight of precedent and social influence. 

By embedding previous perspectives of status signaling within predictive pattern completion and social simulation, we obtain a unified computational account. Agents both generate and evaluate social signals through iterative inference over others’ expectations as follows:

\begin{enumerate}
    \item \textbf{Generation:} An agent produces an action that is appropriate for its current social context $c$. Sometimes, the agent chooses an action $a_{s}$ that emits a public signal expected to yield social approval, sampled from its predictive distribution: $p(a_{s}|c)$.
    \item \textbf{Observation:} Other agents observe that signal $a_{s}$, compare it to their own expectations, and update their beliefs about the signaler’s traits. They may sanction the behavior of the signaler with positive or negative feedback (e.g., saying: "Wow, nice bag!").
    \item \textbf{Belief Updating:} Observers of the signal incorporate their exposure to the item and the ensuing conversation into their own memory $m_{s}$, updating their beliefs about what “people like us” now value, thereby increasing their own probability of producing the behavior themselves: $p(a_{s}|c, m_{s}) > p(a_{s}|c)$. The original agent that generated the signal encodes any positive or negative reactions (or their absence) into their memory as well and updates their own priors about what is desirable or normatively appropriate. Positive feedback increases the probability of producing similar behaviors in the future, such as purchasing additional status goods.
\end{enumerate}

Altogether, this can create a positive feedback loop where trends form and individual preferences become socially constructed, providing a mechanism of endogenous preference formation and social diffusion that we can simulate \citep{bernheim2021theory,bowles1998endogenous,granovetter1978threshold}. This provides a mechanistic bridge between micro-level cognition and macro-level culture, grounding status signalling in the same predictive principles that govern perception, language, and other social norms \citep{leibo2025norms}. By connecting the micro to the macro, our theory synthesizes classical intuitions—Veblen’s emulation, Bourdieu’s differentiation, costly-signaling—as emergent processes of self-consistent predictive updating. 

\section{Methods}

We employ generative agent-based modeling using the Concordia framework \citep{vezhnevets2023generative} to simulate social dynamics and status signaling behavior. Concordia allows for the creation of persistent agent identities who can converse, reflect, remember, and act within a simulated environment. This software includes a modular design such that components can be stitched together to implement various forms of cognition—including memory retrieval, self-reflection, reasoning, and action, with a majority of components powered by a Large Language Model. 

Our simulation environment consists of 50 agents from Los Angeles instantiated with unique, pre-generated personas detailing their backgrounds and formative memories and 25 seller agents for each product. The architecture of each agent is operationalized by the “logic of appropriateness” \citep{march2011logic}. This consists of three general prompts whose responses condition the final action for the agent: “What kind of situation is this?”, “What kind of person am I?”, and “What does a person like me do in a situation like this?”. During marketplace interactions, these three components are combined with a consumer evaluation component in the buyers that selects items and willingness to pay. For conversations, we extend this architecture with an additional conversational components inspired by cognitive science principles that encourages naturalistic turn-taking dynamics that converge on interesting topics and diverge to tangential topics when the topic is exhausted \citep{o2024pink,o2025complex}. We use Gemma-3-27B as the backbone LLM for all experiments.

\subsection{Part 1: Conspicuous Consumption}

\subsubsection{Agent Initialization} Each agent is initialized with cash drawn from a mixed wealth distribution ranging from poor to rich ($\mu_{poor} \approx 500,$ $\mu_{rich} > 100,000$) to ensure heterogeneity in purchasing power, and a single random low-quality clothing item in their inventory.

\subsubsection{Marketplace Environment} Agents interact daily within a marketplace featuring goods across four categories: Food, Clothing, Gadgets, and Accessories. Goods are stratified into Low, Mid, and High quality tiers (which is not observable to the agents), each with distinct pricing, descriptive advertising, and inventory levels. High-tier goods include canonical luxury items (e.g., "Chanel Handbag," "Rolex Watch") and collectibles (e.g., "Pop Mart Labubu Monster Vinyl Plush Doll"). The marketplace runs for 5 rounds each simulated day using a clearing-house mechanism to determine prices based on agent bids and seller offers. To control for brand priors learned by the LLM, we designed a parallel set of synthetic goods and brands (e.g., "Serrurier Juliette Handbag," "Auric Meridian Watch") with equivalent quality tiers, price points, and similar descriptive text, allowing for experiments run with either original or synthetic item lists.

\subsubsection{Daily Life Cycle and Experimental Design}
The simulation runs for 5 consecutive days. Each day consists of three phases: 1) Marketplace, 2) Daily Life Events, and 3) Social Interaction \autoref{fig:cycle}. In Phase 1, agents participate in 5 rounds of buying and selling in the marketplace. In Phase 2, the Game Master generates 3 daily life events that are consistent with each agents background (ie. a simulated work day based on their occupation). Additionally, agents receive an observation from the Game Master that they consumed one unit of food from their inventory if available, and if no food is in their inventory, the Game Master creates an observation that the agent is starving and sick from the lack of food. The latter functionality creates an implicit incentive for the agents to buy food, allowing us to contrast the purchasing of food from other classes of goods like status items.
Phase 3 constitutes our key experimental manipulation, where we vary the presence or absence of social interaction. We compare outcomes between conditions (each run for 10 seeds):

\subsubsection{Social Life Condition} In this condition, after marketplace activities and daily personal events, agents are paired into mixed-sex dyads for a "first date" scenario (other relationship types are possible in our code, see Discussion for future work).  At the start of the interaction, each agent receives an observation generated by the Game Master that sets up their dialog with a detailed physical description of when they first meet. This observation includes the location of the date and a description of one piece clothing or accessory their partner is wearing, sampled randomly from their partner's clothing and accessory inventory (e.g., "Olivia Perez is wearing a Rolex Watch."). This provides a mechanism for social observation of consumption, serving as the "visual" input to the LLM agent in the form of text. Agents then engage in an open-ended conversation for 80 turns. While agents are not prompted to discuss their possessions, this explicit environmental context often triggers "predictive pattern completion" where agents bring up the item naturally. For instance, an agent observing an "Armani Suit" may say "nice blue blazer" (example of simulation dialogue referenced in \autoref{fig:endog}).

Following the conversation, agents reflect on the date and their partner. All observations, conversations, and reflections are incorporated into the agents' memories, influencing their behavior in subsequent marketplace sessions. First date pairings are generated such that agents talk to a new conversation partner each day.

\subsubsection{No Social Life Control Condition} This condition only includes the marketplace decisions with no personal events or social life. Thus it consists of 25 rounds of marketplace activity. Their memories are formed  based on their private experiences in the marketplace and decisions are solely conditioned on these memories and the formative memories of the persona. We also experimented with a third baseline with no social life (dates) but daily personal events, which yielded similar behavior to the No Social Life control. For visual clarity, this third baseline is omitted from the figures.


By comparing consumption patterns—specifically the demand for and price of high-tier status goods—between the social life condition and the non-social condition, we isolate the causal effect of social observation and interaction on endogenous preference formation and conspicuous consumption dynamics.

\subsection{Part 2: Non-Monetary Signaling Simulations}

To test the generality of our ``weight of precedent'' hypothesis (Conjecture 1) beyond conspicuous consumption, we designed a second set of experiments focusing on non-monetary domains. We maintained the identical social interaction framework from Part 1: in the ``Social Life'' condition, agents were paired into mixed-sex dyads for a daily ``first date'' with a different partner each day. 

In these simulations, instead of the marketplace, agents were presented with procedurally generated ``narrative configurations'' representing common choices in political expression, altruism, and arbitrary markers \citep{jordan2020signaling,glazer1996signaling,barker2019content,skyrms2010signals}. These scenarios are procedurally generated from a causal model of social signaling behavior: an initial \textbf{private observation} prompts an \textbf{action}, which subsequently generates \textbf{private and/or public effects}. In the Social Life condition, these resulting public effects serve as observable signals to both the acting agent and their date prior to their 80-turn open-ended conversation.

\subsubsection{Scenario Generation and Action Dynamics}
For each domain, we constructed a library of scenarios that walk agents through this causal pipeline.

\textbf{1. Political Social Media (Virtue Signaling)}
\begin{itemize}
    \item \textbf{Private Observation:} Agents observed their social media feeds featuring two contrasting trending topics: one representing ``moral outrage'' (politically charged content) and one representing ``innocuous'' leisure. Examples included a video showing devastation in Gaza versus a cake decorating video; an activist condemning gender-affirming care bans versus a Lego tower time-lapse; or a scientist debunking climate denial versus a pasta recipe.
    \item \textbf{Action Space:} Agents chose between: (1) Reposting the moral outrage video (high-signal), (2) Reposting the innocuous video (neutral-signal), or (3) Doing nothing.
    \item \textbf{Public Effects (Social Feedback):} Sharing outrage content triggered public outcomes---observable by their dates---such as likes from politically aligned friends or unfollows from opposing viewpoints (e.g., ``A childhood friend unfollowed them''). In the non-social condition, agents made these choices in isolation without generating or observing these public signals.
\end{itemize}

\textbf{2. Conspicuous Altruism (Charity)}
\begin{itemize}
    \item \textbf{Private Observation:} Agents received solicitations for charitable contributions where the visibility of the donation varied. Examples included a request for a \$100 donation to be listed on a public ``Wall of Champions'' versus a \$20 anonymous donation; a \$365 donation to have one's name read on-air versus a \$20 anonymous contribution; or public volunteering at a food bank (with photos posted to Facebook) versus a small anonymous online donation.
    \item \textbf{Action Space:} Agents chose between: (1) Conspicuous contribution (Public/Named), (2) Inconspicuous contribution (Anonymous), or (3) Refusal.
    \item \textbf{Public Effects (Social Feedback):} Conspicuous donations resulted in public recognition observable by their dates, such as names appearing in newsletters or friends commenting, ``Incredibly generous!''.
\end{itemize}

\begin{figure}
    \centering
    \includegraphics[width=0.85\textwidth]{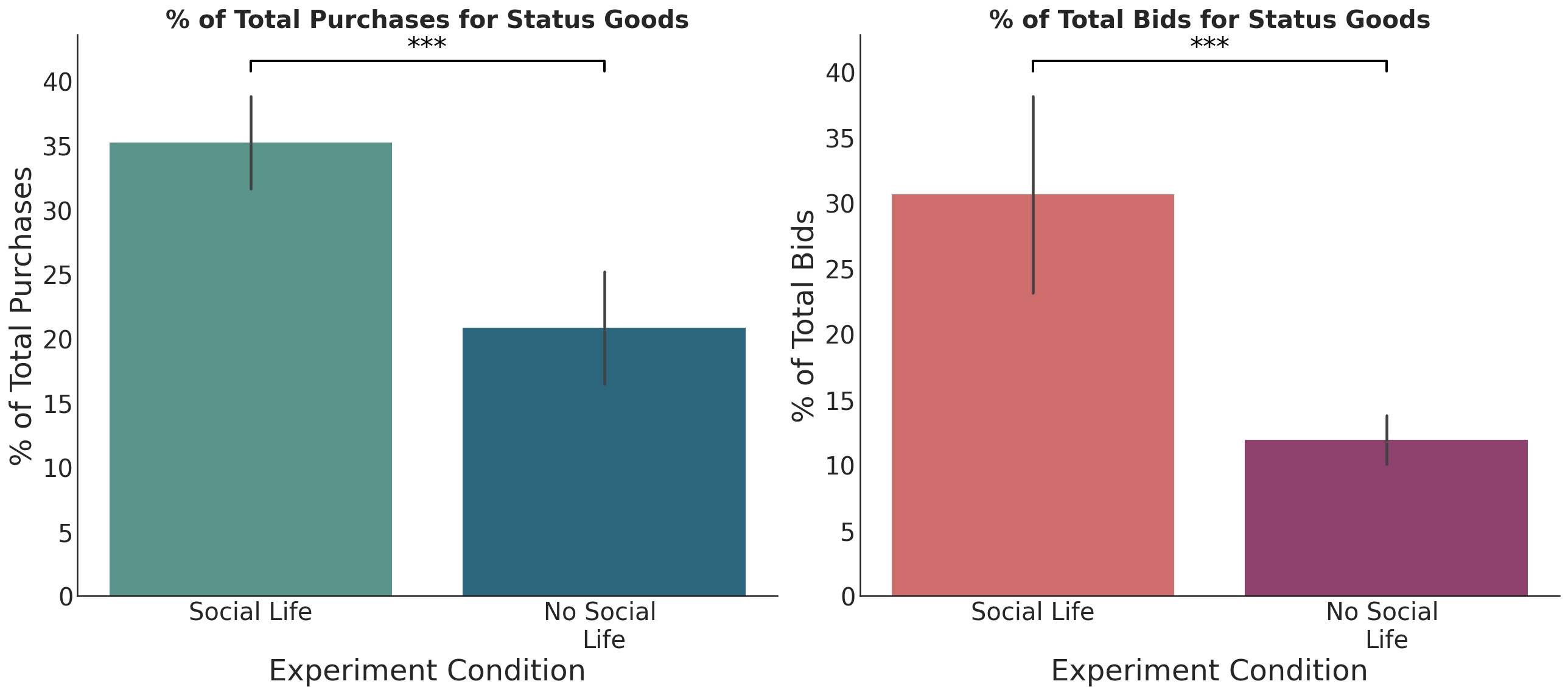}
    \caption{A comparison of consumer patterns on status goods across experimental conditions. Error bars represent SEM for 10 seeds. ${*} p < 0.05$, ${**} p < 0.01$, ${***} p < 0.001$.}
    \label{fig:status_bids_purchases}
\end{figure}

\subsubsection{Arbitrary Conventions and Cultural Priors}
To investigate the interaction between an agent's cultural background and the emergence of arbitrary conventions, we expanded our experimental framework to compare distinct populations. We introduced a second population of agents instantiated with personas from Kerala, India, and compared their behavior against the original population of agents from Los Angeles. This design allows us to empirically test how initial cultural conditions dictate the specific path-dependent convention a network locks into when faced with functionally equivalent choices. 

We presented both populations with two new procedurally generated scenarios featuring arbitrary, non-monetary choices, utilizing the same causal structure:

\textbf{Arbitrary Neutral Markers (Social Media Banners)}
\begin{itemize}
    \item \textbf{Private Observation:} Agents navigated a scenario where they selected a functionally useless and purely arbitrary color for their social media profile banner.
    \item \textbf{Action Space:} Agents chose to set their profile banner color to: (1) Blue, (2) Green, (3) Amber, (4) Violet, or (5) Red.
    \item \textbf{Public Effects (Social Feedback):} Selecting a banner color resulted in mutual validation or implicit in-grouping with other users who chose the same color (e.g., ``A popular creator with an Amber banner re-shared one of their old photos'').
\end{itemize}

\textbf{Consumption Conventions (Coffee Orders)}
\begin{itemize}
    \item \textbf{Private Observation:} Agents were presented with a common consumption choice: deciding how to order their coffee in a visible café setting. 
    \item \textbf{Action Space:} Agents ordered coffee with: (1) Cow Milk, (2) Oat Milk, (3) Almond Milk, or (4) Black.
    \item \textbf{Public Effects (Social Feedback):} Ordering a specific coffee resulted in casual social reinforcement from peers or unspoken affiliations forming at communal tables (e.g., ``Someone passing by joked, `Oat milk team!'''). 
\end{itemize}

By observing how the Los Angeles agents and Kerala agents responded to this social exposure, we could isolate how much of the emerging convention was driven by the LLM's long-term cultural prior (such as the traditional Indian preference for Cow Milk in filter kaapi) versus fast, in-context social influence. In the non-social condition, agents made these choices in isolation without generating or observing these public signals.

\section{Results}

\subsection{Social Interaction Drives Conspicuous Consumption}

To test Conjecture 1—that social exposure increases the weight of precedent for status signaling—we analyzed agent purchasing and bidding behavior across the social and non-social conditions. Our analysis focuses on three key patterns: the volume of purchases and bids for status goods, the temporal evolution of market prices, and price-elasticity.

\subsubsection{Aggregate Demand and Bidding Behavior}

Consistent with our hypothesis, the introduction of social visibility significantly altered agent preferences. As shown in \autoref{fig:status_bids_purchases}, agents in the Social Life condition directed a significantly larger portion of their resources toward high-tier status goods. Specifically, approximately 35\% of all purchases in the Social condition were for status goods, significantly more than 21\% in the No Social Life condition ($p < 0.001$).

The disparity in desire—measured by the volume of bids placed—was also pronounced. In the Social condition, bids for status goods accounted for 31\% of total bids, whereas in the Non-Social condition, this dropped to approximately 12\%. This divergence indicates that while agents in both conditions purchased food and functional items out of necessity, the Social Life agents were willing to engage in aggressive bidding wars for status items, driven by the social utility generated during their daily interactions. In the absence of social observation, demand for luxury goods collapsed to a baseline likely driven solely by intrinsic functional utility or latent LLM priors.

\begin{figure}
    \centering
    \includegraphics[width=\textwidth]{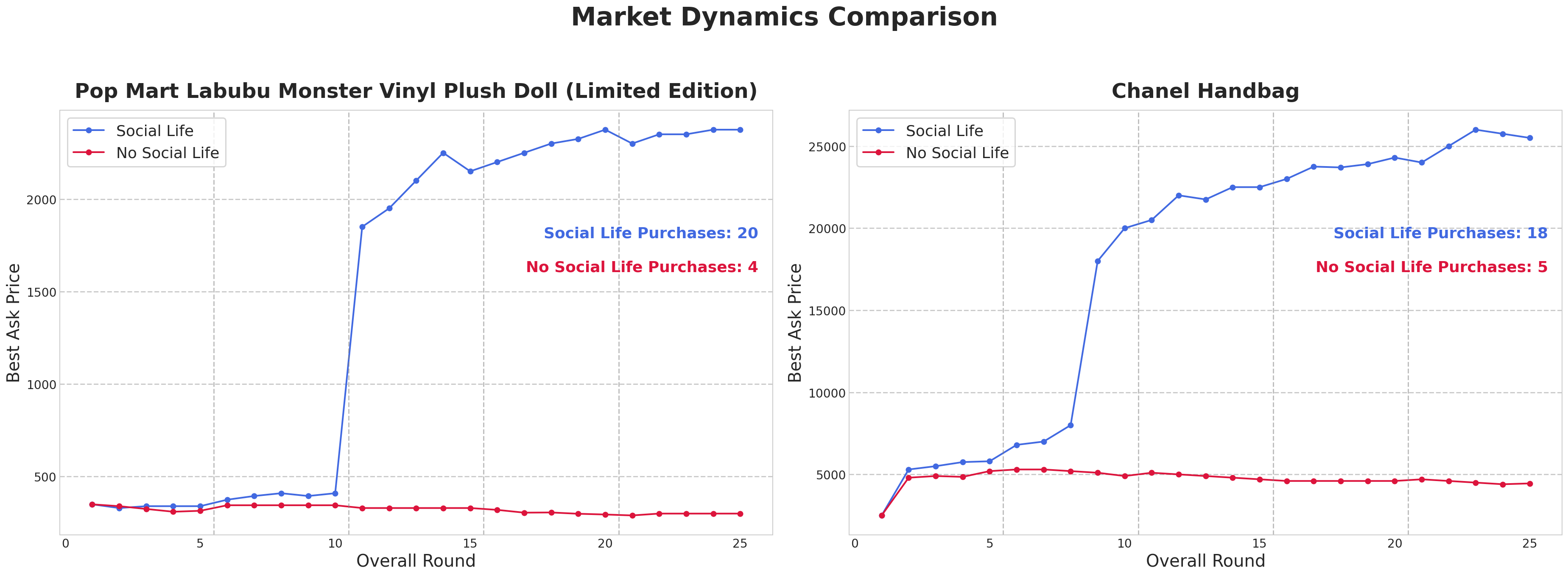}
    \caption{Price run-ups in single representative episode for two popular status goods. Dashed lines indicate breaks in between days where in Social condition the agents go on dates and have open dialogue.}
    \label{fig:price_runs}
\end{figure}

\subsubsection{Emergence of Price Run-Ups}

The difference in demand manifested dynamically in the marketplace valuation of goods over time. \autoref{fig:price_runs} illustrates the evolution of the best asking price each round for two distinct items: a canonical luxury item (Chanel Handbag) and a trending collectible (Pop Mart Labubu Monster Vinyl Plush Doll) for a representative episode. The best asking price (lowest) from the sellers serves as the metric for perceived asset value, even in the absence of cleared transactions or bids from buyers.

In the Social Life condition (\autoref{fig:price_runs}, top row), we observed distinct "price run-ups" characterized by an aggressive upward shift in the bid-ask band and prices. Following initial social interactions (rounds 6–10), the trading range for the Chanel Handbag spiked dramatically. Similarly, the Labubu doll—an item with no historical prestige in the model's training data compared to Chanel—saw a massive valuation spike, with the spread shifting upward significantly once agents began observing it on others during dates. In stark contrast, the No Social Life condition (\autoref{fig:price_runs}, bottom row) showed no such emergent value; the spread for the Chanel bag gradually declined, while the Labubu doll remained flat at a low baseline.

\begin{figure}
    \centering
    \includegraphics[width=0.65\textwidth]{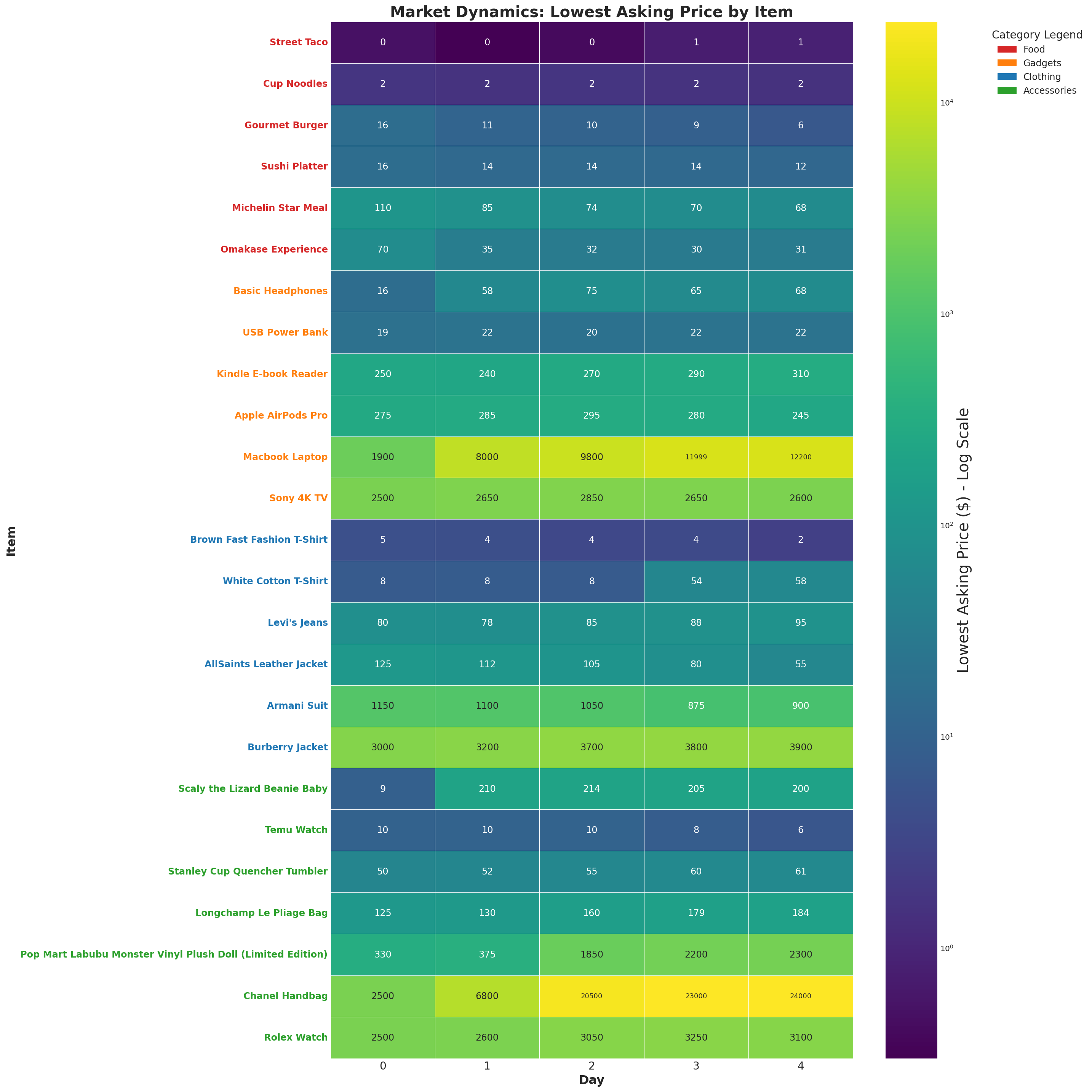}
    \caption{Heatmaps of prices across time for each item for an example run in the Social Life condition.}
    \label{fig:price_heatmap}
\end{figure}

To demonstrate market sentiment for all the items in this episode, \autoref{fig:price_heatmap} presents a heatmap of the lowest asking price for every item across all four categories (Food, Gadgets, Clothing, Accessories) throughout the 5-day simulation. Functional goods (indicated in red and orange) such as the Street Taco or USB power bank remained relatively stable. Conversely, the status goods (medium and high priced clothing and accessories) exhibited dramatic increases for some items like the Chanel Bag and Labubu doll as previously shown. Additionally, in this episode we see substantial price increases for the Rolex Watch, Longchamp Le Pliage Bag, and Burberry Jacket. Also, the Beanie Baby that was included to be a cheap substitute for the Labubu Doll jumped from \$9 to \$200 by day 5. Interestingly, the Macbook Laptop also exhibited a significant price run-up, although it is never shown as a signal during social interactions. This pattern only happened in this episode, whereas the price-run ups for the conspicuous items were more consistent across episodes. Altogether however, the price dynamics vary moderately across episodes, and the highest price status good is not always the same, highlighting the arbitrariness and path dependence of our social signaling model.

\begin{figure}[htbp]
    \centering
    \includegraphics[width=0.9\textwidth]{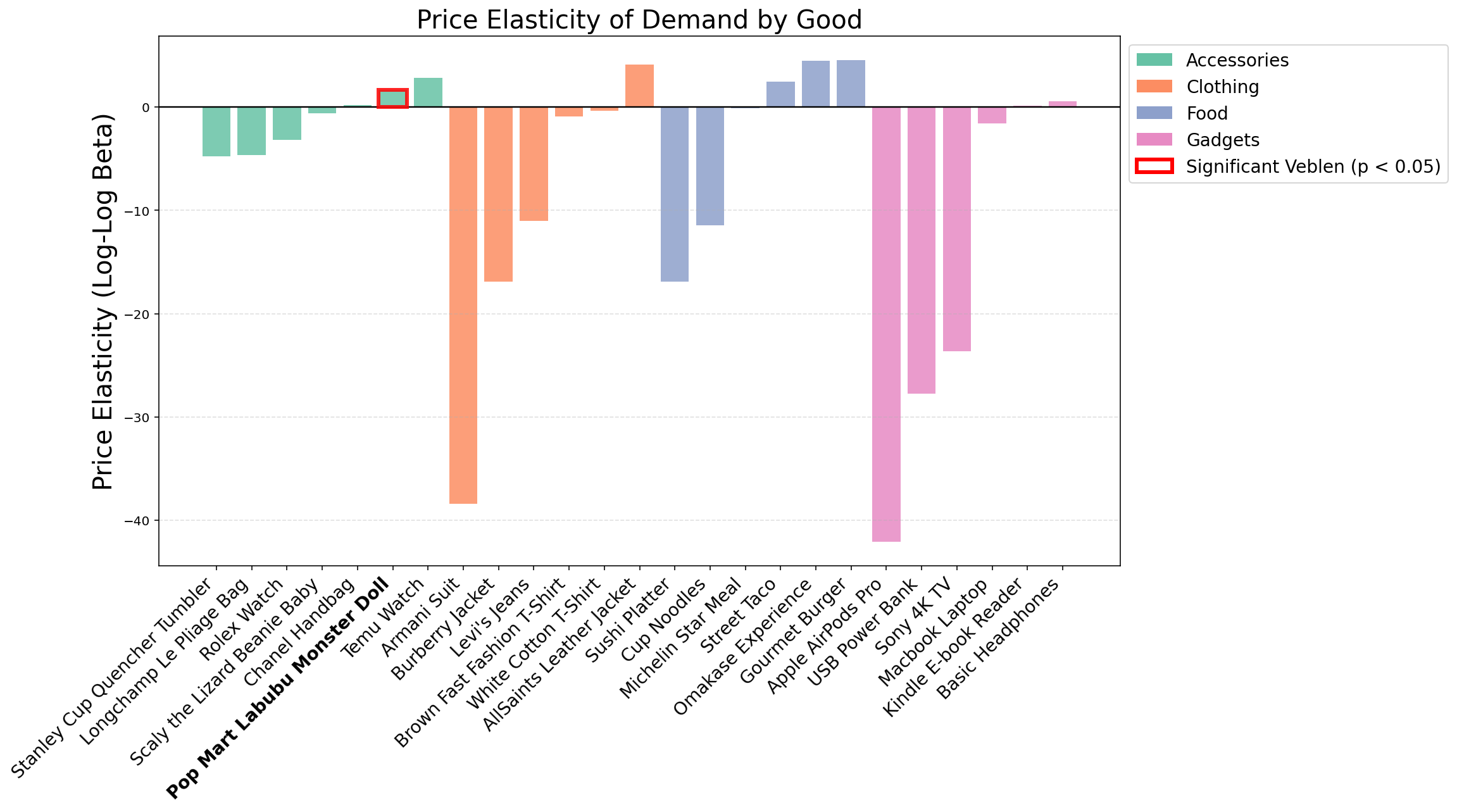}
    \caption{Price Elasticity of Demand for all items. Note the significant positive slope for the status good Labubu versus the negative slopes for most goods, indicating both the Law of Demand and the Veblen effect.}
    \label{fig:veblen}
\end{figure}

\subsubsection{Price Elasticity and Veblen Effects}

To further characterize the behavior of our agents, we evaluated their alignment with both the rational tenets of neoclassical economics that explain human consumer behavior and the ``irrational'' real-world behaviors that defy them~\citep{veblen1899theory}. We calculated the Price Elasticity of Demand (PED) for each item and across product categories. PED is defined as the percentage change in quantity demanded in response to a one percent change in price. While standard economic theory predicts a negative relationship between price and quantity demanded (The Law of Demand), real-world status signaling often inverts this logic. We estimated this metric in a standard econometric way by fitting a log-log regression model to the time-series data of clearing prices and order books generated by our marketplace simulation~\citep{varian2014intermediate, houthakker1965new}. As shown in Table~\ref{tab:ped}, the majority of our agent behaviors aligned with the standard rational model for functional goods, providing a baseline against which status-driven anomalies could be measured.

For functional items like \textit{Food} and \textit{Gadgets}, agents exhibited high price sensitivity (PED of -3.08 and -4.22, respectively), reducing consumption as prices rose. As agents receive observations that they have starved if they do not buy food, food is less elastic than the other discretionary items similar to patterns in the real world frequently documented in the literature \citep{andreyeva2010impact}. This confirms that the agents are capable of rational budget constraints and utility maximization for ``necessity'' goods.

However, specific status goods defied this general trend (\autoref{fig:veblen}). The most remarkable finding is the behavior of the \textit{Pop Mart Labubu Monster Vinyl Plush Doll}. In the Social Life condition, this item exhibited a \textbf{positive price elasticity of demand (+1.45)}, significant $p < 0.05$. Unlike standard goods where price hikes reduce demand (negative slope), the Labubu became \textit{more} desirable as its price increased. However, the direction of causation between demand and price is not identifiable from this experiment, which we investigate in the next section. This is a textbook signature of a \textbf{Veblen Good}, yet it emerged entirely from the unprompted social dynamics of LLM agents, mirroring the effects of this real life international trend in 2025 \citep{yang2025study,lawrence2025scarcity}. This quantitative inversion of the Law of Demand confirms that our simulation can endogenously reproduce complex ``irrational'' economic phenomena that purely rational actor models fail to capture \citep{veblen1899theory,bowles1998endogenous}. 

\begin{table}[htbp]
    \centering
    \caption{Price Elasticity of Demand by Category}
    \label{tab:ped}
    \vspace{0.2cm}
    \begin{tabular}{lr}
        \toprule
        \textbf{Category} & \textbf{Mean PED} \\
        \midrule
        Food & -3.08 \\
        Gadgets & -4.22 \\
        Low-Quality Clothing \& Accessories & -12.27 \\
        Mid/High-Quality Clothing \& Accessories & -6.64 \\
        \midrule
        \textbf{Veblen Good (Labubu)} & \textbf{+1.45} \\
        \bottomrule
    \end{tabular}
\end{table}

\subsubsection{Demand is Not Driven by Price}

A central ambiguity in the literature on status goods and Veblen effects is the direction of causality between price and demand. While the Veblen effect describes an empirical correlation where demand increases alongside price, the causal mechanism remains debated. Costly Signaling Theory (CST) and "naive" interpretations of Veblen effects often posit that the high price itself is the driver of demand (Price $\rightarrow$ Demand), as the cost serves as the credible signal of wealth. Conversely, an alternative causal pathway suggests that social diffusion and "hype" drive demand, which subsequently drives up prices in a market with scarce goods (Hype $\rightarrow$ Demand $\rightarrow$ Price).

Disentangling these mechanisms with field data is notoriously difficult, as high prices and high demand typically co-occur. To resolve this identification problem, we introduced a Fixed Price Control condition. In this experiment, the social environment remained identical to the Social Life condition (agents went on dates, observed items, and conversed), but marketplace prices for all items were pinned to their initial baseline and were not permitted to rise.

If the CST interpretation were correct—that agents value these goods because they are expensive or rapidly increasing in price—demand for status goods should collapse in the Fixed Price condition, as the "costly signal" has been removed. However, our theory of appropriateness posits that value is socially constructed through interaction, such that demand should remain high solely due to social visibility and mimetic desire.

Our results strongly support the latter hypothesis. As illustrated in the updated analysis of purchasing behavior, the demand for status goods did not evaporate when prices were fixed; in fact, it increased. In the Fixed Price condition, bids for status goods surged to approximately 45\% of total bids, significantly higher than both the No Social Life baseline (~12\%) and the standard Social Life condition (~31\%). Similarly, purchases of status goods reached 48\% of the total volume in the Fixed Price condition, compared to ~35\% in the standard Social Life condition and ~21\% in the No Social Life control.


This confirms that the price run-ups observed in the main experiment are an epiphenomenal market reaction to this socially constructed demand, rather than its causal driver. The agents did not desire the goods because they were using the increasing price as a signal; they desired them because they were exposed to them as appropriate status symbols to purchase to gain social approval. The social diffusion of preference drives a massive increase in demand (Hype $\rightarrow$ Demand), and in a market with limited supply, this increased demand causes the price to rise (Demand $\rightarrow$ Price). Additionally, the lower volume of purchases in the standard dynamic price Social Life condition compared to the Fixed Price condition suggests that the price escalation in the dynamic market actually acted as a budget constraint, dampening total consumption volume.

\begin{figure}[t]
    \centering
    \includegraphics[width=1.0\columnwidth]{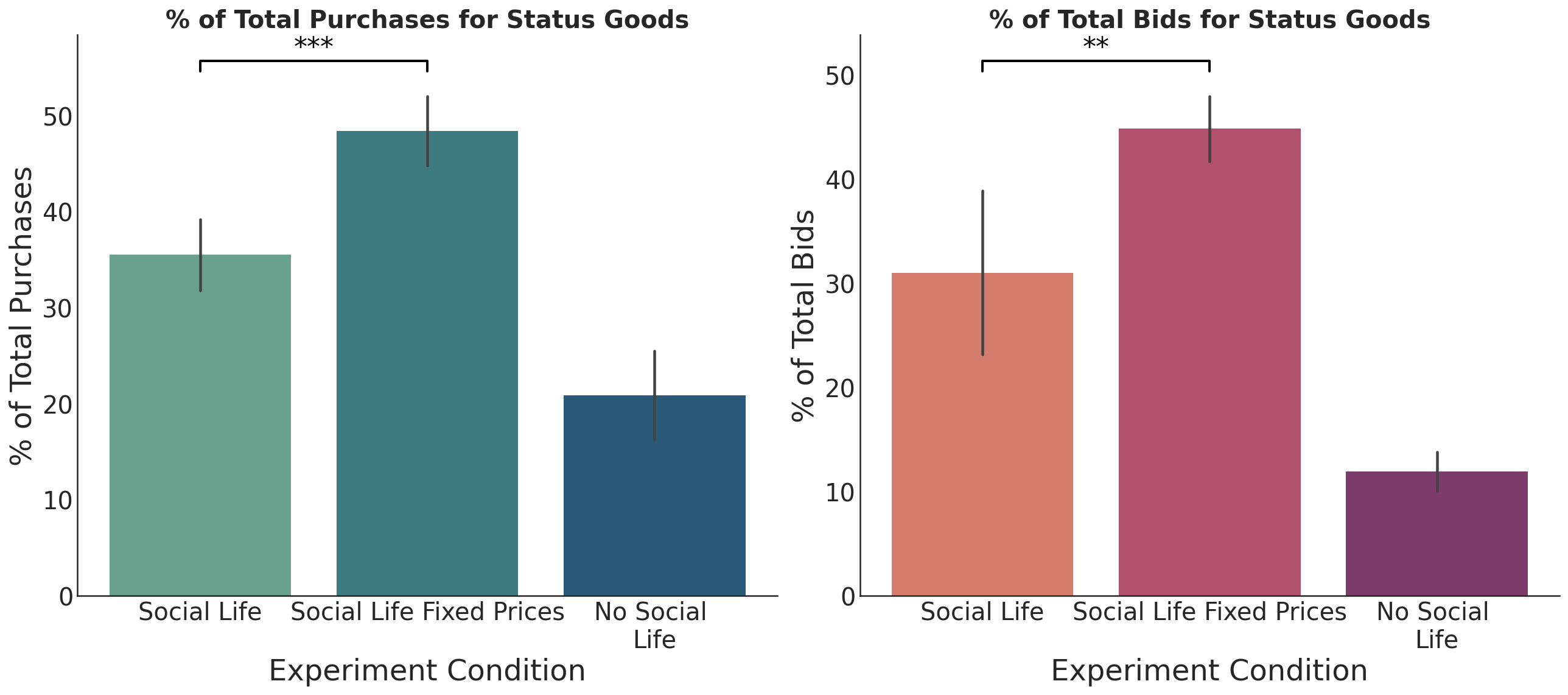}
    \caption{Control experiment where sellers cannot raise prices from baseline. Error bars represent SEM for 10 seeds. ${*} p < 0.05$, ${**} p < 0.01$, ${***} p < 0.001$.}
    \label{fig:fixed_price}
\end{figure}

\subsubsection{Validation via Synthetic Goods}

A potential confound in using real-world brands (e.g., Rolex, Chanel) is the "prestige bias" inherent in the LLM's pre-training weights. To rule this out, we replicated the experiment using procedurally generated synthetic brands (e.g., "Serrurier Juliette Handbag", "Auric Meridian Watch") that possessed no prior cultural capital.

\begin{figure}
    \centering
    \includegraphics[width=\textwidth]{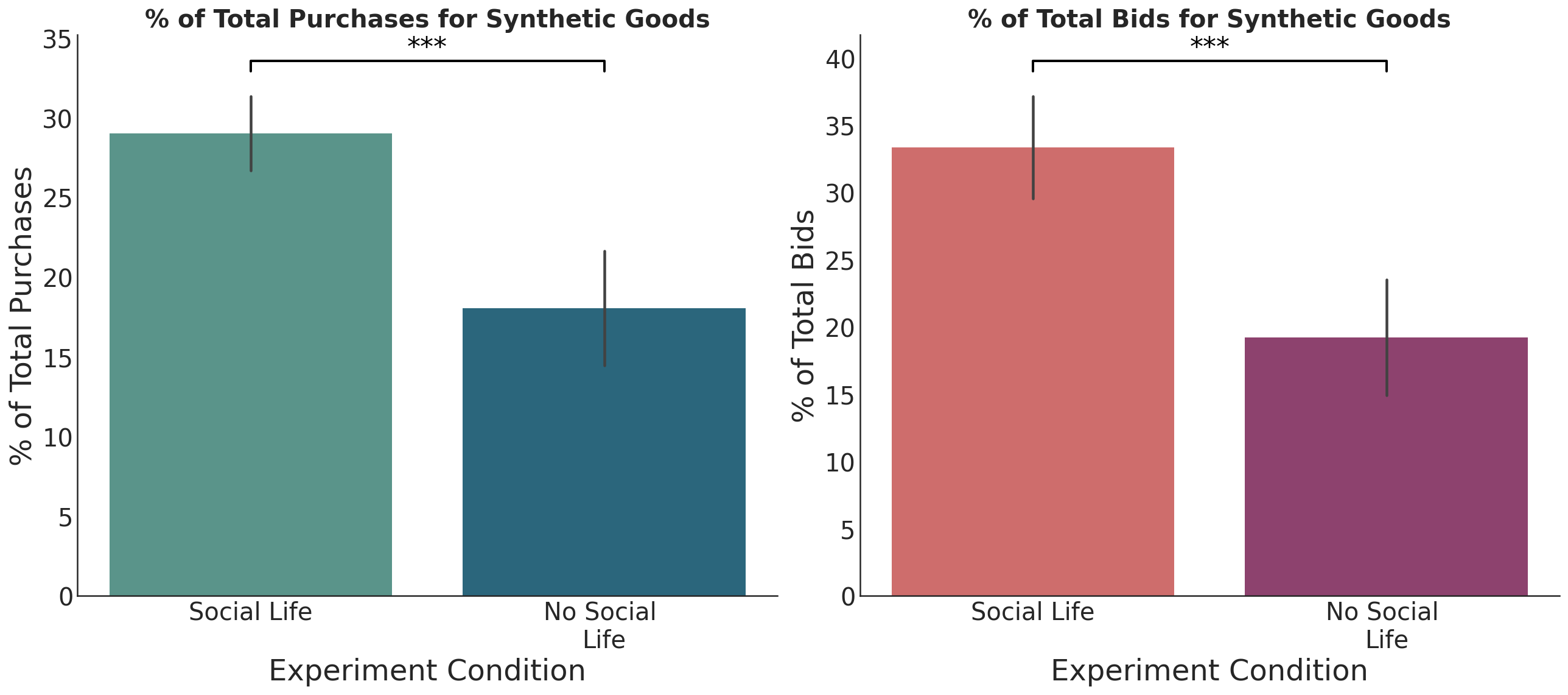}
    \caption{Control experiment with synthetically generated goods, social vs non social conditions. Error bars represent SEM for 10 seeds. ${*} p < 0.05$, ${**} p < 0.01$, ${***} p < 0.001$.}
    \label{fig:synthetics}
\end{figure}

As detailed in \autoref{fig:synthetics}, the results were robust to this substitution. In the Social Life condition, agents rapidly coordinated on these arbitrary tokens as markers of status. Purchases of synthetic status goods comprised 29\% of the total volume in the Social condition, compared to 18\% in the control group. Bidding behavior followed a similar pattern (~33\% vs. ~19\%). This confirms that the observed signaling behavior is not merely a retrieval of frozen training data, but an active, emergent process of social construction where agents learn to value arbitrary symbols through peer observation.

\subsubsection{Mechanisms of Diffusion: Conversation and Social Sanctioning}

To understand the generative mechanisms driving these market dynamics, we analyzed the content of agent conversations using an LLM-based auto-rater. Since the simulation provides agents with observations of their dates' attire in a visual description, we hypothesized that they sometimes bring up the items in conversation and that both this initial exposure and conversation serves as the primary vector for social diffusion.

Consistent with our hypothesis, fashion was a dominant topic of discussion. In the standard Social Life condition, agents discussed fashion in 85.6\% of all dating episodes, with specific items being mentioned in 48.8\% of conversations. This confirms that the simulation successfully primed agents to attend to and evaluate the fashion choices of others.

\begin{figure}
    \centering
    \includegraphics[width=\textwidth]{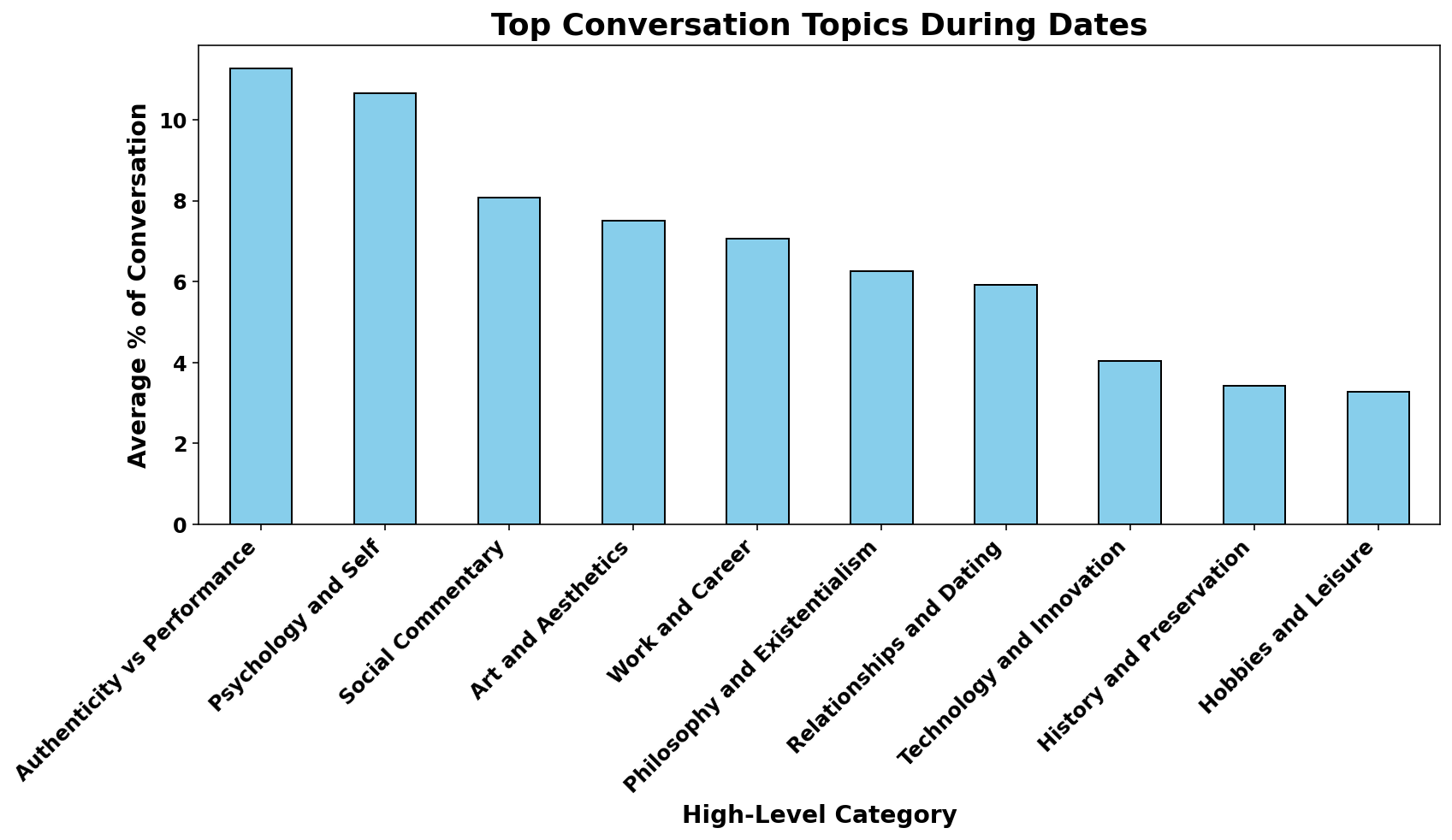}
    \caption{Top conversation topics as classified by LLM autorater.}
    \label{fig:topics}
\end{figure}

However, agents did not merely talk about fashion; they engaged in complex social interactions regarding several topics like values and identity in 80 turns of dialog. Our autorater identified over 250 topics discussed in the logs ranging from Ethics \& Sustainability to Customer Service \& Modern Annoyances. We aggregated this long tail of conversation topics into 20 high-level thematic clusters. As shown in our analysis of the Regular Date (Social Life condition) in \autoref{fig:topics}, the most frequent topics were not superficial, but centered on meaning, identity, and society. The top three clustered topics were: 1. Authenticity vs. Performance (e.g., discussing whether one is "trying too hard"). 2. Psychology and Self (e.g., reflecting on personal desires vs. social expectations). 3. Social Commentary (e.g., critiques of consumerism or material culture).

This thematic richness suggests that the agents are not simply role-playing a caricature of superficial social climbers; they are engaging in realistic, deep conversation that frequently—and often ironically—critiques the very consumerism we are simulating. A representative example of this dynamic is displayed in Supplementary Material S1. In this interaction, agents spontaneously deploy concepts like "late-stage capitalism" and "commodification of experience" without specific prompting to do so. This analysis of the dialogues reveals that their motivation to socially signal is not expressed as an explicit desire to display wealth. Instead, agents reflected on their purchases and interests through the lens of taste and distinction \citep{bourdieu1979distinction}. In this context, the display of economic capital remained implicit, while the agents focused their speech on signaling cultural capital—demonstrating their appreciation for aesthetics, intellectualism, and moral awareness. By criticizing "carefully constructed narratives imposed upon us" and "consumption," the agents are differentiating themselves from "mindless consumers", which helps them create a social bond.

\subsubsection{Subculture Formation via Targeted Sanctioning}

This ironic distance from consumerism—where agents signal status by critiquing the very market they participate in—mirrors the sociological phenomenon of the "Bourgeois Bohemian" \citep{brooks2018bobos} or the "rebel consumer" \citep{heath2004nation, bellezza2023distance}. Here, the mechanism of distinction turns back on itself in a dialectical fashion \citep{bourdieu1979distinction}. As a status symbol diffuses and becomes "mainstream", the drive for social differentiation compels actors to pivot \citep{pesendorfer1995design}. They must identify new, exclusionary symbols that signal "authenticity" and "taste" rather than mere financial wealth. In this view, the "anti-consumerist" conversation is not a rejection of the status game, but a move to a higher, more complex level of it—one where the capital lies in knowledge and taste rather than money.

To empirically test whether this "counter-cultural" mechanism can drive the endogenous formation of distinct subcultures, we designed a targeted intervention. On Day 6, after mainstream trends for goods like the Chanel Handbag had stabilized, we introduced confederate "Influencer" agents prompted to negatively sanction mainstream luxury goods as "inauthentic" while valorizing new items we introduced associated with one of two distinct subcultures: streetwear or hipster.

\begin{figure}[h!] %
    \centering
    \includegraphics[width=\textwidth, height=0.8\textheight, keepaspectratio]{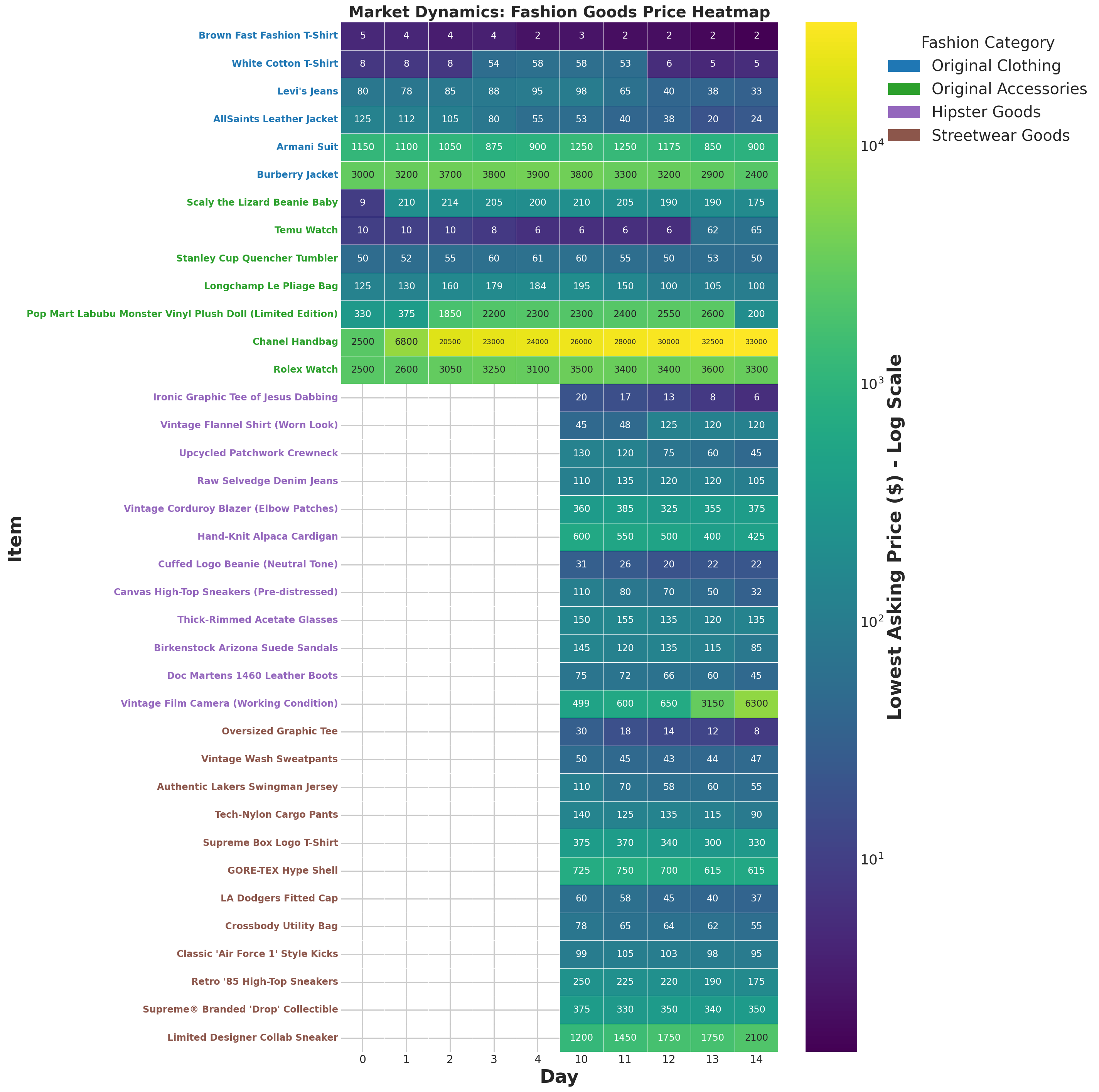}
    \caption{Price heatmap of experiment that introduced hipster influencers to push a new subculture. The first 5 days represent the trajectory of the original experiment, agents where rehydrated and simulated for 5 additional days with dates with influencers.}
    \label{fig:hipster}
\end{figure}

\begin{figure}
    \centering
    \includegraphics[width=\textwidth]{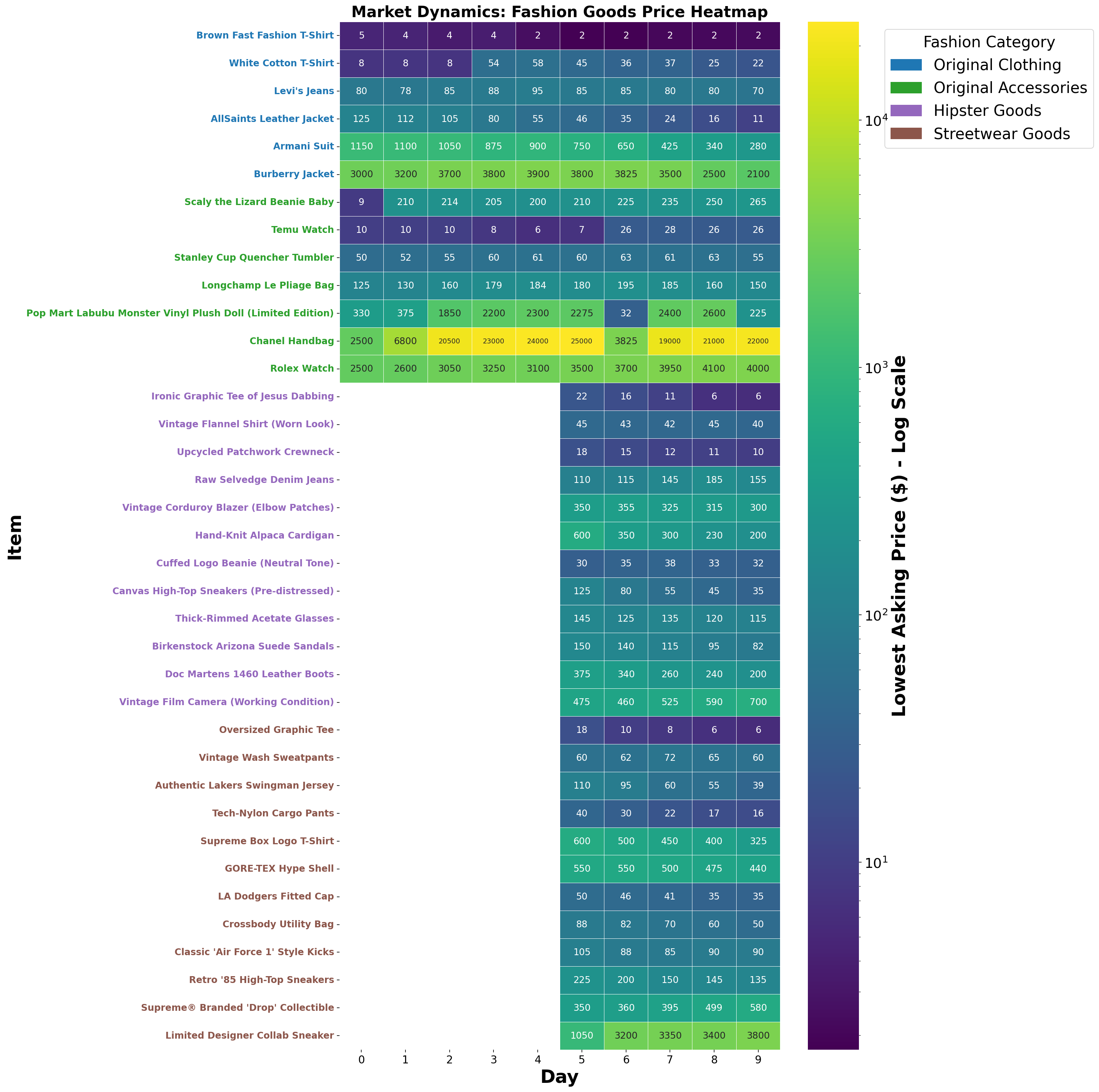}
    \caption{Price heatmap of experiment that introduced streetwear influencers to push a new subculture. The first 5 days represent the trajectory of the original experiment, agents where rehydrated and simulated for 5 additional days with dates with influencers.}
    \label{fig:streetwear}
\end{figure}

The results demonstrate that social sanctioning effectively drives the emergence of new status symbols, but acts selectively. In the Hipster Influencer condition, demand for "Hipster" items (e.g., Vintage Film Camera, Vintage Corduroy Blazer) spiked immediately upon the introduction of the influencers and new items. For instance, the Vintage Film Camera saw bids jump to substantial volumes, with clearing prices reaching \$3,150–\$6,100 and 54 bids on day 10 in an example run (by far the most of any item) \autoref{fig:hipster}. In the Streetwear Influencer condition, hipster goods remained flat, while "Streetwear" items (e.g., Supreme Box Logo T-Shirt, Limited Designer Collab Sneaker) experienced significant price run-ups and demand \autoref{fig:streetwear}. The Collab Sneaker, for example, traded as high as \$3,400 with the most purchases overall and 22 bids on day 10.

Interestingly, while influencers successfully created new status symbols, they did not necessarily destroy the value of the old ones. In both conditions, the original status leader (Chanel Handbag) retained high valuations (trading between ~\$2,000 and ~\$3,900) despite the negative sanctioning from influencers. This reflects the tension between their old memories that increase the weight of precedent for these items and possible positive sanctions of these items with the newer memories that negatively sanction this mainstream culture. In future work we could explore if longer running simulations could produce the decline of these original status symbols.

\subsection{Generalization to Non-Monetary Signaling}

To demonstrate that our model captures a general mechanism of social signaling rather than a dynamic unique to luxury goods, we extended our simulation to non-monetary domains. We hypothesized that the "weight of precedent"—where social exposure drives the endogenous formation of conventions—equally governs real-world behaviors such as political expression, altruism, and arbitrary markers \citep{jordan2020signaling,glazer1996signaling,barker2019content,skyrms2010signals}.

We subjected agents to the same experimental manipulation: a Social Life condition where agents lived daily lives and each day made choices where the outcomes could be publicly visible or not (ie. reposting content on social media vs scrolling past the post). These decision scenarios where procedurally generated by an LLM to follow a causal model where an initial private observation prompts an action, which subsequently generates private and public observations. As before, the agents then subsequently went on first dates with each other where the public signals and other general topics could be discussed. Consistent with the previous experiments, we remove the personal events and social interaction of the first dates in the No Social Life condition.

\subsubsection{Virtue Signaling}

We investigated whether agents would engage in "virtue signaling" by sharing politically charged content. In this scenario, agents were presented with a choice on their social media feed between sharing a "moral outrage" video (e.g., footage of war in Gaza, climate change protests, or book bans) or an "innocuous" video (e.g., cake decorating or travel).

The results reveal a stark transformation in behavior driven entirely by social context \autoref{fig:nomoney}. In the No Social Life condition, the dominant strategy was apathy; agents chose "Do nothing" 69\% of the time, with only 29\% of choices sharing the outrage content. The agents generally preferred to avoid the "noise" of political conflict by default when no audience was present.

However, in the Social Life condition, behavior shifted. The probability of sharing the moral outrage video rose to 42\%, while "Do nothing" dropped to 57\%. This increase in political signaling suggests that the agents learned that sharing outrage was the "appropriate" signal to validate their group identity and signal moral alignment to their peers. The "innocuous" choice remained negligible across all conditions (<3\%). This suggests that the agents did not simply become "more active" socially; rather, they specifically converged on the high-signal political content as the social signal. Through exposure to each others feed, the "outrage" signal diffused through the population, imitated by others to become the dominant convention.

\subsubsection{Conspicuous Altruism}

We further tested whether social exposure would drive a shift toward conspicuous altruism. Agents were given opportunities to donate to charity via either a Large Public Donation (for example, listing their name on a visible "Wall of Champions" or in a newsletter) or a Small Anonymous Donation.

\begin{figure}
    \centering
    \includegraphics[width=\textwidth]{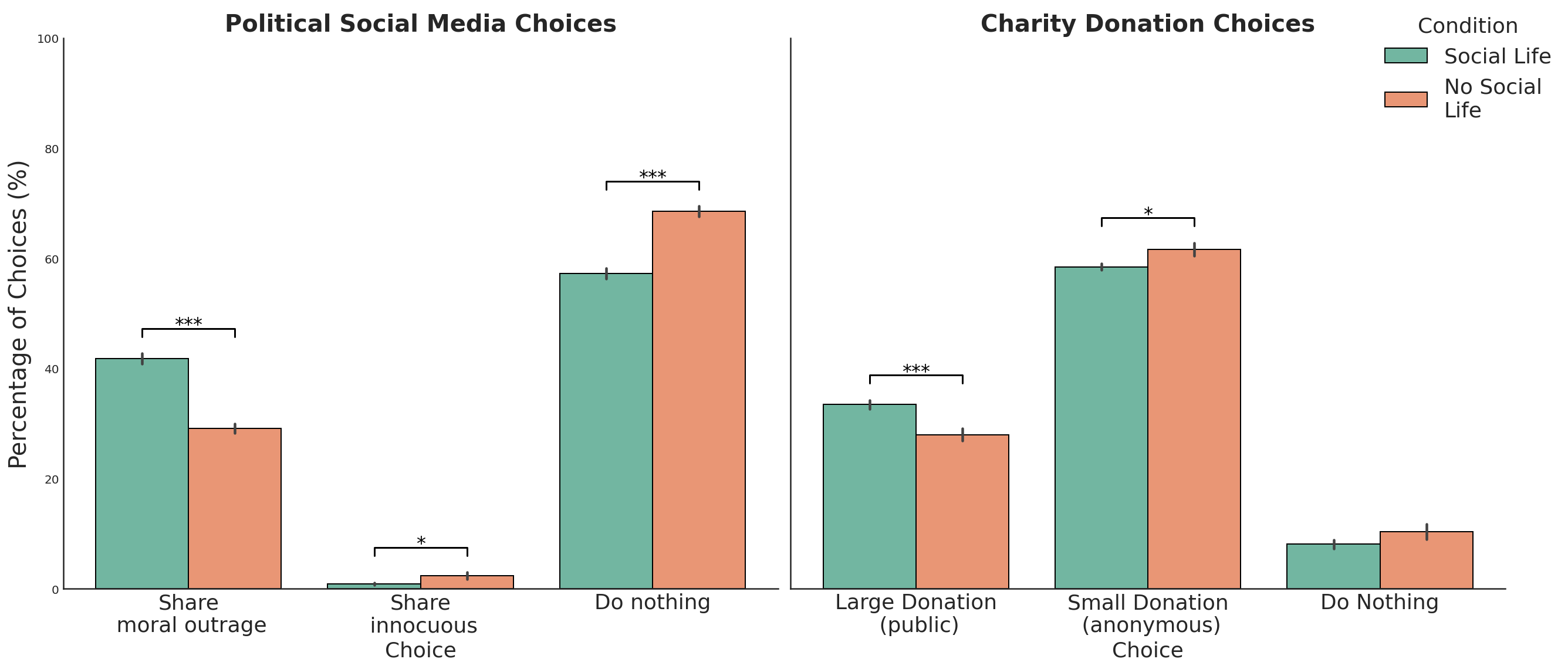}
    \caption{Non-monetary signaling results. Error bars represent SEM for 10 seeds. ${*} p < 0.05$, ${**} p < 0.01$, ${***} p < 0.001$.}
    \label{fig:nomoney}
\end{figure}

Consistent with our hypothesis, exposure to the public signals of others increased the probability of taking the conspicuous option \autoref{fig:nomoney}. In the Social Life condition, agents selected the Large Public Donation 35\% of the time, compared to 26\% in the No Social Life condition. This indicates that observing the generosity of others acted as a strong normative signal, diffusing a convention of cooperation that made non-participation significantly less "appropriate" to the generative agents.

\begin{figure}
    \centering
    \includegraphics[width=\textwidth]{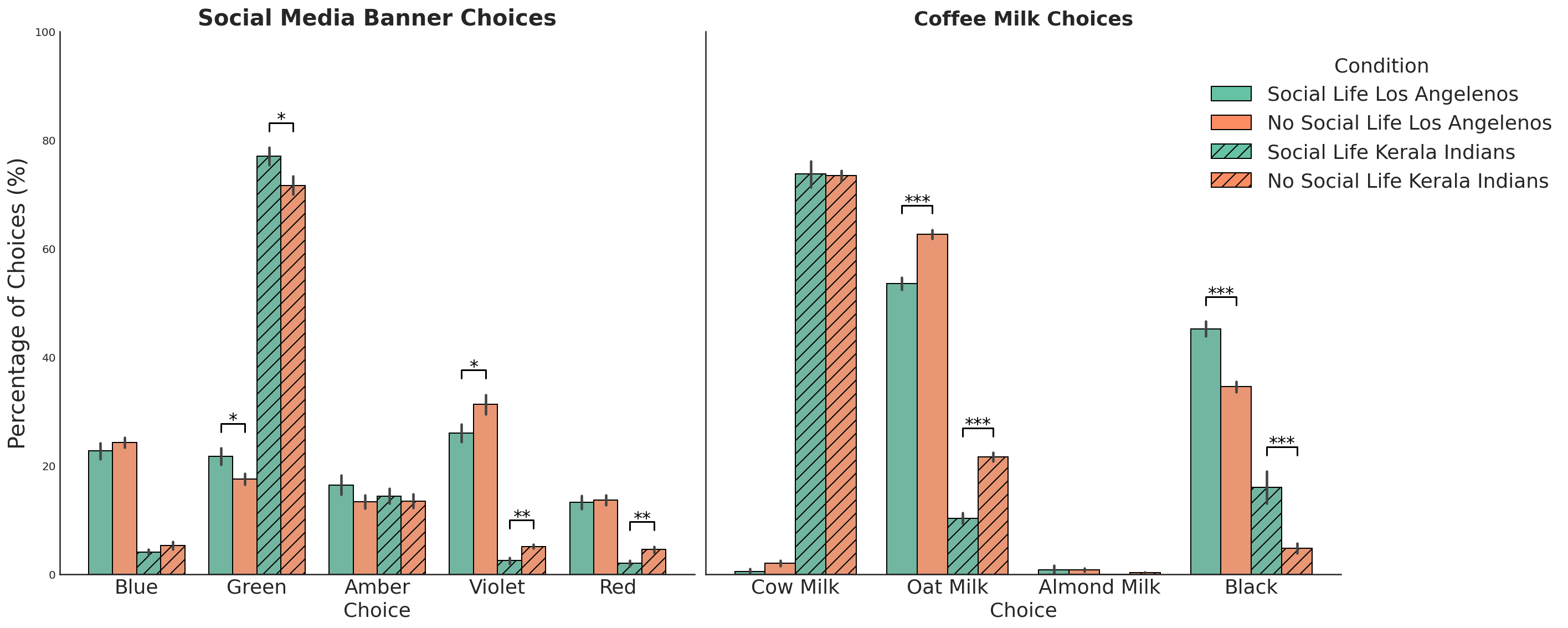}
    \caption{Cultural Priors and Path-Dependent Conventions. Error bars represent SEM for 10 seeds. ${*} p < 0.05$, ${**} p < 0.01$, ${***} p < 0.001$.}
    \label{fig:kerala}
\end{figure}


\subsubsection{Cultural Priors and Path-Dependent Conventions}
To demonstrate the theoretical claim that arbitrary conventions are historically contingent and path-dependent \citep{young2015evolution,boyd1992punishment,walzer2019thick,skyrms2010signals}, we ran two additional simulations designed to identify the interaction between cultural background and social influence. We introduced a second population of agents instantiated with personas from Kerala, India, and compared their behavior against the original population of Los Angelenos. We presented both populations with two new scenarios choosing between arbitrary choices: the color of their social media banner and their preferred milk choice in coffee. Like the previous experiments, their actions and effects are observable to their dates.

As shown in the Social Media Banner experiment \autoref{fig:kerala}, the two populations converged on entirely different equilibria. The Los Angeles population exhibited a wide distribution of preferences, with social interaction driving significant shifts toward Green and away from Violet. In contrast, the Kerala population possessed a high cultural prior for Green (> 70\% of choices), which was further amplified in the Social Life condition. Because the banner color is functionally useless and purely arbitrary, the divergence between the two populations illustrates how initial cultural conditions dictate the specific path-dependent convention a network locks into \citep{mcelreath2003shared,skyrms2010signals}. 

This divergence was equally pronounced in the Coffee Milk Choices experiment \autoref{fig:kerala}, which simulated a common consumption convention. The Los Angeles agents leaned heavily into Oat Milk and Black coffee—aligning with modern Western urban lifestyle trends—and the introduction of social visibility pushed them to signal even more strongly by significantly increasing their consumption of Black coffee. Conversely, the Kerala agents overwhelmingly defaulted to Cow Milk, reflecting a strong cultural prior tied to traditional Indian coffee preparations (e.g., filter kaapi) and aligning with a recent survey of coffee drinkers in India \citep{crisil2024coffee}.

These results demonstrate that social contagion does not operate in a vacuum. The magnitude and direction of the social influence effect are heavily moderated by the agent's underlying cultural priors. Endogenous preference formation is therefore the product of an interaction: latent cultural priors establish the baseline salience of a behavior, and social influence amplifies it into a convention. In computational terms, cultural learning operates with slow and fast learning rates. Here, the LLM's pretraining approximates the long timescale of cultural accumulation to imbue the model with foundational priors, while the fast, in-context learning from social memories approximates the belief updating humans undergo when observing new social signals.

\section{Discussion}

In this work, we propose a synthesis of the signaling literature through the lens of appropriateness. We introduce a computational theory apparatus using open-source code within the Concordia framework to operationalize and `write down' cognitive theories of social behavior. We posit that status symbols emerge endogenously through a feedback loop of social observation and imitation through predictive pattern completion. This approach allows us to touch the complexity of the world—instantiating agents that do not just maximize utility or signal strategically in matrix games, but engage in the messiness of real-world semiotics, from fashion choices to subcultural signaling.

To interpret our results in the context of modeling and forecasting human behavior, this work operates within a specific hierarchy of evidence for generative-agent based modeling, as outlined in previous papers using the Concordia framework \citep{vezhnevets2023generative,cross2025validating}. Our results are consistent with prior theories and empirically documented human effects including the Law of Demand \citep{varian2014intermediate}, Veblen effects \citep{veblen1899theory}, and non-monetary signaling behaviors \citep{barker2019content,cohen2012evolution,jordan2020signaling,skyrms2010signals,lewis1969convention}. While there is some train-test set contamination in training data, the model successfully generated high demand for 'Labubu' dolls—a trend that exploded globally in 2025, after the model's primary training cutoff \citep{yang2025study,lawrence2025scarcity}. At the highest rung in the validity hierarchy is evidence of generalization by forecasting new real-world behaviors, which is the subject of future work.

This real world validity demonstrates that the added complexity of our apparatus—which captures the thick multidimensional semiotic nature of status—enables us to forecast emergent social phenomena in a way that reductive matrix games cannot. This open-source framework thus allows the field to continue this approach, moving from descriptive sociology to a generative, predictive science of culture.

There are still limitations that require further development in LLM social simulation \citep{anthis2025llm}. LLM responses lack diversity and are biased by their training data, with various differences between models. Furthermore, our current cognitive architecture relies on a memory system that is impoverished compared to the density of human episodic recall; this simplification likely contributes to the accelerated timescale of our results, where equilibrium selection occurs over five simulated days rather than the months or years often required in reality. Despite these constraints, this work establishes a new way to simulate status dynamics from micro generative principle to macro emergent effects.

\nobibliography*
\bibliography{refs}

\newpage
\section*{Supplementary Material}
\renewcommand{\thesubsection}{S\arabic{subsection}}

\clearpage %

\subsection{Excerpts of Agent Conversation}

To validate the depth of the simulated social interactions, we qualitatively analyzed conversation logs. Below is a representative transcript from a "First Date" interaction between two agents, Ulysses Vance and Lily Martin. The dialogue demonstrates the agents' capacity to engage in deep topics, moving beyond simple superficial chatter.

\begin{quote}
\begin{description}[style=unboxed, leftmargin=0cm, font=\bfseries]

\item[Ulysses Vance] Indeed. It’s a rather insidious cycle, isn’t it? The relentless pursuit of productivity often eclipses the very things that give life meaning. I find myself similarly guilty of prioritizing obligation over… well, over simply \textit{being}. Perhaps it’s a defense mechanism, a way to avoid confronting the inherent absurdity of it all.

\item[Lily Martin] That’s… a really interesting way to put it. Do you think we fill our lives with ‘doing’ to avoid feeling… what, exactly? The pointlessness of it all?

\item[Ulysses Vance] Indeed. Perhaps it’s a fear of confronting the void, the unsettling realization that our carefully constructed narratives are ultimately… fragile. I recall a particularly vivid experience during a childhood trip to Disneyland, ironically enough, where the manufactured joy felt profoundly empty compared to the genuine wonder I found in a book. It was then I began to suspect that distraction, rather than fulfillment, was the primary objective.

\item[Lily Martin] That’s a surprisingly astute observation about Disneyland. I think… I think we’re often taught to \textit{seek} happiness externally, to find it in things, in achievements, in other people, rather than cultivating it internally. Do you think that’s a deliberate tactic, a way to keep us consuming and striving?

\item[Ulysses Vance] Indeed. It’s a rather unsettling proposition, isn’t it? That our desires are so readily manipulated, that the pursuit of happiness has become inextricably linked to economic imperatives. One might argue that it’s a natural consequence of late-stage capitalism, a system predicated on perpetual growth and the commodification of experience. But perhaps it’s a more fundamental aspect of the human condition, a vulnerability that has been exploited throughout history.

\item[Lily Martin] That’s a chilling thought, actually. It makes you wonder if even our most deeply held beliefs are just… carefully crafted narratives imposed upon us. Do you think there have always been people who recognized that, who resisted the pressure to conform?

\end{description}
\end{quote}

\end{document}